\documentclass[a4paper,11pt]{article}
             
\usepackage{jinstpub} 

\usepackage{multirow}
\usepackage{CJKutf8} 
\usepackage{cleveref}
\usepackage{graphicx}      
\usepackage{subcaption}    
\usepackage[english]{babel}
\title{\boldmath A 64-Channel Precision Time-to-Digital Converter with average 4.77 ps RMS Implemented in a 28 nm FPGA}

\author[a]{Zehong Liang,}
\author[a,1]{Xiongbo Yan,}
\author[a,1]{Zhe Ning,}
\author[a]{Jun Hu,}
\author[a]{Xiaoshan Jiang,}
\author[a]{Yunhua Sun,}
\author[a]{Weiyan Pan,}
\author[a]{Jingbo Ye,}

\note{Corresponding author.}

\affiliation[a]{Institute of High Energy Physics, University of Chinese Academy of Sciences}

\emailAdd{yanxb@ihep.ac.cn}
\emailAdd{ningzhe@ihep.ac.cn}

\abstract{We have developed a Time-to-Digital Converter (TDC) application in a Xilinx Kintex-7 Field Programmable Gate Array (FPGA).  This TDC, based on the Tapped-Delay Line (TDL) and Wave Union A (WU-A) techniques, achieves an independent time measurement on 32-channel rising edges and 32-channel falling edges. The average time resolution or the Least Significant Bit (LSB) of the 64 channels is measured to be 3 ps level, with an average root mean square (RMS) precision of 4.77 ps, and a maximum RMS below 8 ps. We also propose an online processing scheme that handles the bubble issues caused by clock region skew.}


\proceeding{December 23, 2024}

\begin{document}

\selectlanguage{english}
\maketitle
\flushbottom

\section{INTRODUCTION}
\label{sec:intro}
Time-to-Digital Converters (TDC) play a pivotal role in a multitude of applications\cite{RN1}, including high-energy physics\cite{RN2,RN3}, nuclear medicine imaging technology\cite{RN4,RN5}, and radar technology\cite{RN6,RN7}. Nowadays the Full Width at Half Maxima (FWHM) of particle detector can reach 300 ps and below\cite{RN8,RN9,RN10,RN11}, with the resolution and precision requirements of TDC are also increasingly improved.

Through extensive research and development by the academic community over the years, TDCs based on FPGA have made substantial progress. FPGA TDCs are capable of achieving time resolutions in less than 10 picoseconds (ps)\cite{RN12,RN13,RN14,RN15,RN16,RN17}, showing potential that is on par with Application-Specific Integrated Circuit (ASIC) based TDCs., making FPGA-based TDCs highly attractive for various applications.

Most high-precision TDCs based on FPGA achieve time measurement through the implementation of coarse and fine timestamps. The coarse timestamp is generated by coarse counter and the fine timestamp is generated by fine time structure. The coarse counter structure is employed to extend the range of the fine timestamp, and the fine time structure is tasked with time interpolation. Based on the method of implementing the fine time structure, FPGA TDCs can be essentially categorized into several types: Phased Clocks, Tapped-Delay Lines (TDL), Differential TDC, and Pulse Shrinking\cite{RN18}. 

The fine time structure for TDL TDCs typically consists of a series of delay units that are similar in size and connected end-to-end to flip-flops. This interconnected chain of units is known as tapped-delay line. The temporal resolution capability of TDL structure is derived from the delay time of the fundamental delay unit, with the common approach being the use of dedicated fast carry logic units within FPGAs, namely CARRY4.

CARRY4 is composed of four carry logic units, and for FPGAs fabricated with 28nm process technology, the average propagation delay of each logic unit is approximately 15 ps\cite{RN19}. This is clearly insufficient for a high-precision TDC at the picosecond level. To surmount the resolution constraints imposed by the excessive propagation delay of the logic units in TDL TDCs, there are two commonly used approaches.

The first approach is known as the Wave Union\cite{RN20}. The Wave Union scheme was proposed in 2006 and is based on the principle of multi-measurement, which is divided into two variants: Wave Union A and Wave Union B. The principle of the Wave Union A scheme is to achieve multiple measurements within a single clock cycle, thereby obtaining high-precision time measurement results in a short period of time. However, this approach corresponds to a more complex design of delay chains and encoders, which may present certain challenges in implementation. On the other hand, the Wave Union B scheme realizes multiple measurements across multiple clock cycles, which reduces the design complexity but sacrifices measurement efficiency and is also subject to the limitations of the stability of the oscillator\cite{RN21}.

The second approach is a Multi-TDL scheme. The principle of enhancing measurement accuracy in this scheme also relies on multi-measurement, but it differs from the Wave Union scheme. The multi-measurement of this scheme is accomplished collectively by multiple delay chains\cite{RN22}. This scheme enables rapid measurement while maintaining a simple structure of delay chains. The trade-off is that the consumption of resources increases exponentially with the number of measurements within a single clock cycle.

As the manufacturing process of FPGAs continues to evolve, FPGAs based on 28nm and even 20nm processes are becoming increasingly popular. The delay time of TDL delay units is becoming shorter. Consequently, the skew and jitter in the TDL sampling clock path and the path from TDL to the D flip-flop can no longer be disregarded. Coupled with the characteristic of CARRY4's carry looks ahead\cite{RN23}, these factors lead to the actual acquired tap sequence being a non-standard thermometer code. It is similar to a thermometer code, but it exhibits logic state bits that are not entirely "0" or entirely "1" near the transition edges, known as "bubbles". As the FPGA manufacturing process continues to advance, the problem of bubbles becomes more severe, especially for TDCs based on the Wave Union A scheme with multi-edge waveforms, posing an even greater challenge.

Due to the high resource consumption of multi-TDLs scheme, the Wave Union scheme holds greater potential for the design of multi-channel TDCs compared to the multi-TDL scheme. The Wave Union B scheme can meet the requirements for relatively high precision and time resolution without consuming a large amount of resources, but at the cost of a significant dead time, which can span several or even dozens of sampling clock cycles\cite{RN21}. 

The Wave Union A TDC typically features a smaller dead time. However, it comes with a more severe bubble problem and a more complex design of the delay chain and encoder. Bubbles are prone to confusion with multiple transition edges, making them indistinguishable. Convenient encoding schemes such ones-counter encoder\cite{RN22} is not feasible as well. A new encoding method known as multi-edge decomposition encoder\cite{RN24} has resolved this problem. It eliminates the need for offline reordering of the TDL chain through software and does not require individual bubble corrections along each transition edge. Additionally, it does not introduce any extra dead time, which facilitates the implementation of multi-channel Wave Union A TDC with multiple edges.

In this report, a 4-edge WU-A TDC had implemented based on the multi-edge decomposition encoder within a 28 nm FPGA. A solution was proposed for severe bubble problem based on the tap swapping method, allowing the utilization of FPGA space affected by severe bubbles to achieve a multi-channel TDC. A two-channel TDC was initially validated and then duplicated according to actual requirements to form 32 channels of rising edge TDC and 32 channels of falling edge TDC. Measurement results indicate that the realized TDC is capable of achieving an average accuracy of less than 5 ps.

\section{TDC ARCHITECTURE}
\label{sec:architecture}
\begin{figure}[htbp]
\centering 
\includegraphics[width=0.8\linewidth]{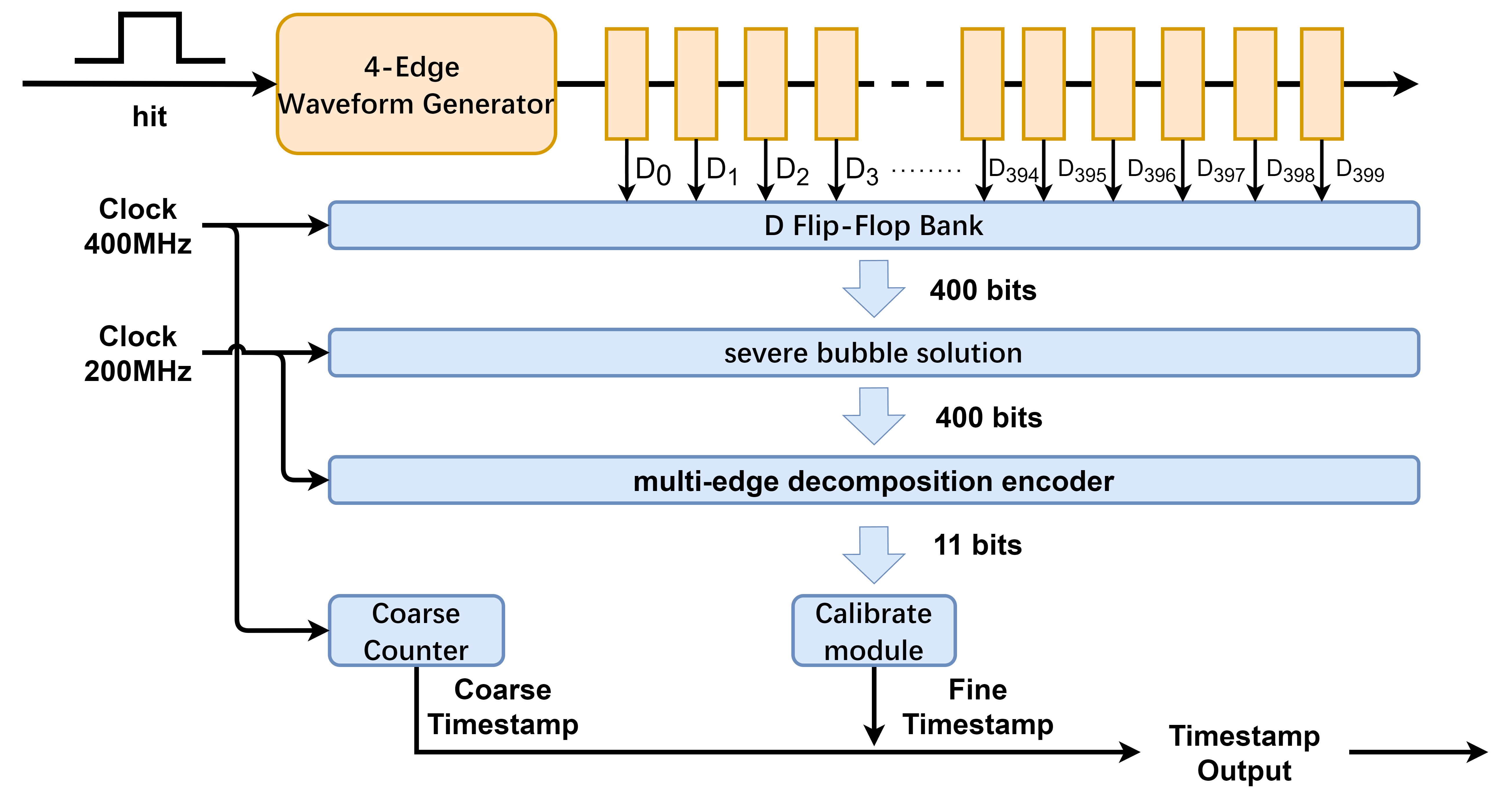}
\caption{\label{fig:1} Structure of the 4-edge TDL TDC.}
\end{figure}

The structure of 4-edge TDC is shown in figure \ref{fig:1}. The Time-to-Digital Converter (TDC) is composed of several components, including a 4-edge waveform generator, a 400 bits Tapped Delay Line (TDL), a D-type flip-flop bank, a module of severe bubble solution, a multi-edge decomposition encoder, a coarse counter, and a calibration module. The coarse counter generates coarse timing information, while the TDL operates at the same clock frequency as the coarse counter to ensure that the fine timestamps can a complete length for time interpolation.

The 4-edge waveform generator preserves time information while converting the sampling signal into a 4-edge waveform. When the waveform enters the TDL, the time information is transformed into the position of the waveform within the TDL. The D-type flip-flop bank reads out the status of the TDL. The severe bubble solution module and multi-edge decomposition encoder form the encoding structure of the TDC, converting the readout from the D-type flip-flop bank into binary encoding. The encoding result represents the sum of the positions of each transition within the 4-edge waveform on the TDL, which is positively correlated with the delay time within the TDL.

Due to the significant nonlinearity of the delay units within the TDL, a calibration module is necessary to individually calibrate the binary encoding results.

The 4-edge waveform generator preserves time information while converting the sampling signal into a 4-edge waveform. When the waveform enters the TDL, the time information is transformed into the position of the waveform within the TDL. The D-type flip-flop bank reads out the status of the TDL. The severe bubble solution module and multi-edge decomposition encoder form the encoding structure of the TDC, converting the readout from the D-type flip-flop bank into binary encoding. The encoding result represents the sum of the positions of each transition within the 4-edge waveform on the TDL, which is positively correlated with the delay time within the TDL.

Due to the significant nonlinearity of the delay units within the TDL, a calibration module is necessary to individually calibrate the binary encoding results.

\section{FPGA IMPLEMENTATION}
\label{sec:implementation}
\subsection{Wave Generator and TDL}
In Xilinx Kintex-7 series FPGAs, each slice contains a dedicated fast carry chain known as CARRY4, which comprises four multiplexers (MUXCY). The delay of each MUXCY is about 15 ps, and the D flip-flops adjacent to CARRY4 can be connected through short routing. Dedicated channels also facilitating rapid cascading between CARRY4. Therefore, CARRY4 can serve as high-quality taps for TDL. 

The length of the TDL is determined by the length of waveform generator and the frequency of sampling clock. The waveform generator needs to be initialized at the starting point of the TDL before the arrival of sampling signal. Therefore, a shorter waveform generator can result in a shorter TDL. The TDL length needs to be slightly longer than one sampling clock cycle, so high-frequency clocks can shorten the TDL length. However, the jitter of high-frequency clocks can also affect measurement accuracy \cite{RN25}. Given that the TDL in this work has up to 400 taps, a lower sampling clock frequency can be appropriately selected.

\begin{figure}[htbp]
\centering 
\includegraphics[width=0.7\linewidth]{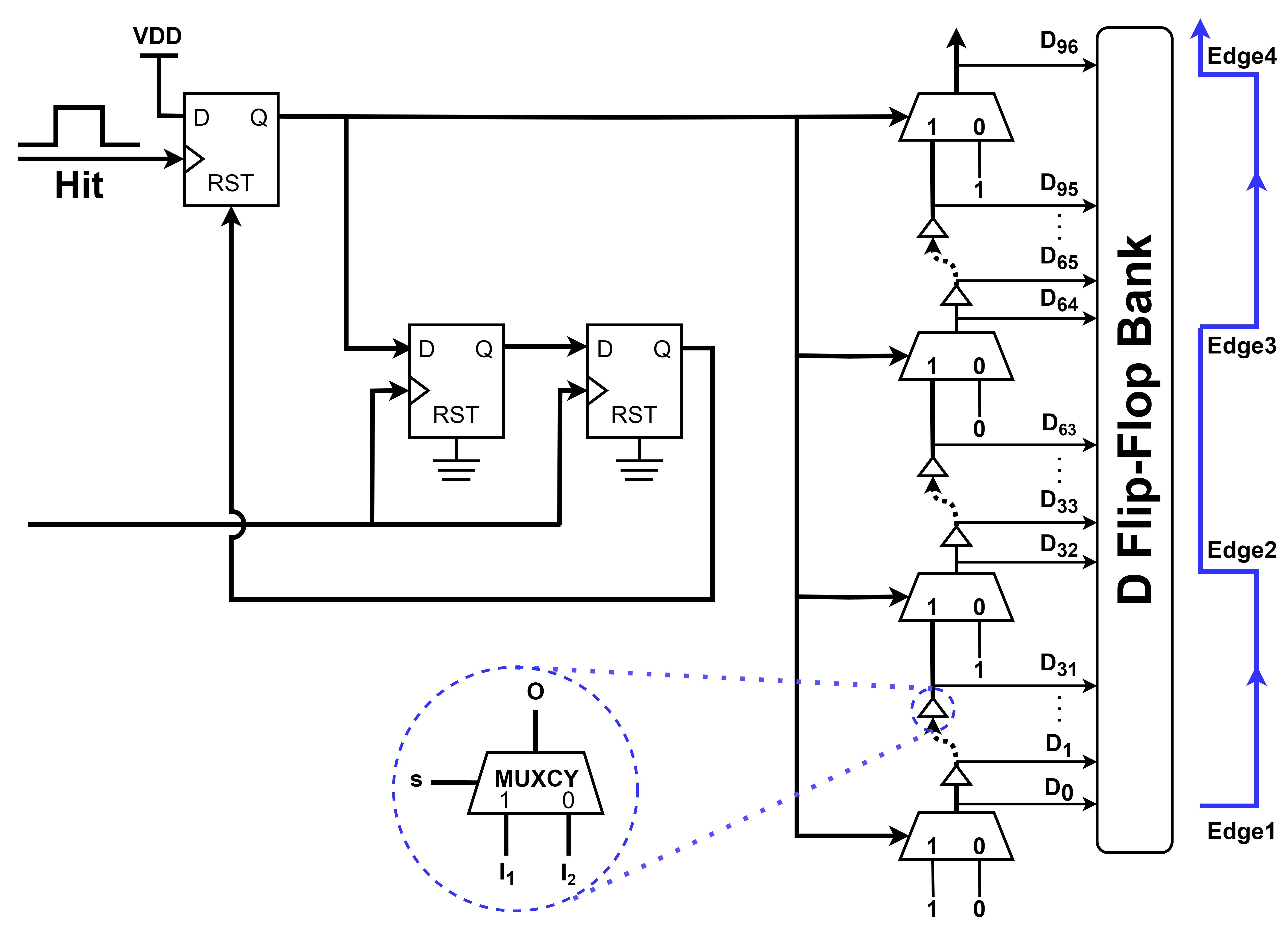}
\caption{\label{fig:2} Structure of the Wave Generator.}
\end{figure}

In this work, TDL is constructed by cascading a total of 100 CARRY4, corresponding to 400 MUXCY outputs. The Wave Generator occupies the first 96 positions, and the delay line occupies the subsequent 304 positions, corresponding to 400 D flip-flop outputs, with a sampling clock of 400 MHz. The structure of the Wave Generator is shown in figure \ref{fig:2}. Three D flip-flops form a stretcher, with an elongation duration of 1-2 sampling clock cycles. The stretched signal is fanned out into 4 paths, each controlling the MUXCYs at positions 0, 32, 64, and 96. When no sampling signal is triggered, the CARRY4 outputs generate a static 4-edge waveform on D flip-flops. Upon the triggering of the sampling signal, the waveform is released and propagates through the TDL. 

Due to the influence of sampling signals arriving during the reset of the stretcher module on the arrangement of the scale within the calibration module, the sampling signals arriving at the time of the stretcher module reset are shielded. The stretcher introduces a total of 3 clock cycles of dead time, which is 3*2.5 = 7.5 ns, with a maximum of 2 clock cycles for stretching and 1 clock cycle for resetting.

\subsection{Online De-bubble solution and encoding scheme}
\begin{figure}[htbp]
\centering 
\includegraphics[width=0.7\linewidth]{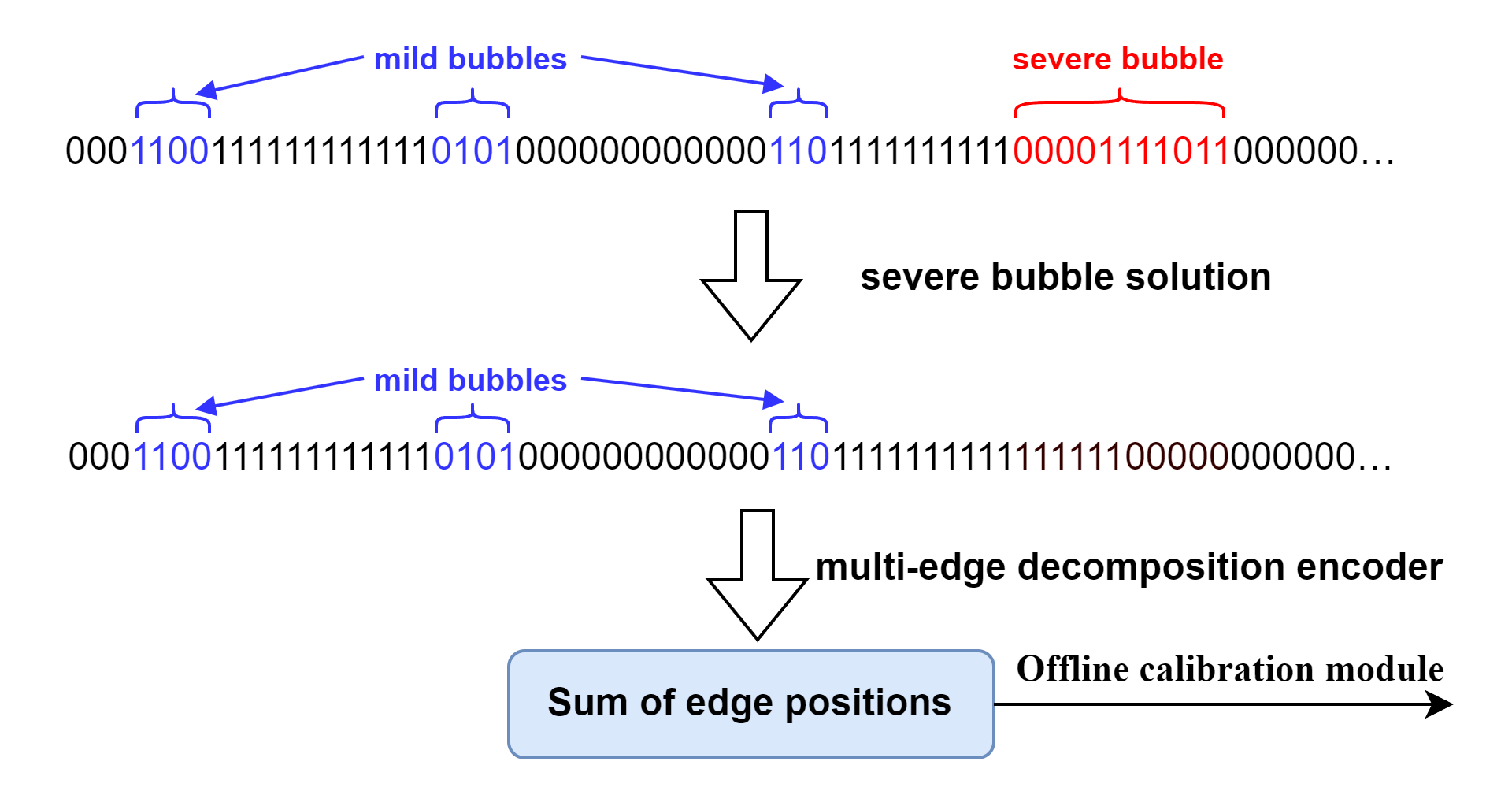}
\caption{\label{fig:3} The overall flow of de-bubble process.}
\end{figure}

The original bit sequence obtained from the D flip-flop is not a standard thermometer code, it is interfered by bubbles and becomes garbled as shown in the example figure \ref{fig:3}. There are two different types of bubbles in the original bit sequence, one of which is severe bubble with a width greater than 10, located at the boundary of two clock regions and caused by clock skew between clock domains. Another type is mile bubble with a width less than 8, which can be located at any position on the tapped delay line, due to the FPGA clock skew, the carry looks ahead in Xilinx devices CARRY4 and so on.

The multi-edge decomposition encoder can perform de-bubble and encoding of multi-edge TDL online, with the encoded output being the sum of edge positions. One-hot code is no longer needed and can be ignored directly, thus requiring less resources and encoding time. However, the distance between edges needs to be large enough to ensure that edges can be recognized without being confused with bubbles. Adapting to severe bubbles will result in larger edge distances, which will lead to more D flip-flop usage, more clock domains, and more severe bubbles. Therefore, it is considered to handle severe bubble first and then perform multi-edge decomposition encoding.

\begin{figure}[htbp]
\centering 
\includegraphics[width=0.8\linewidth]{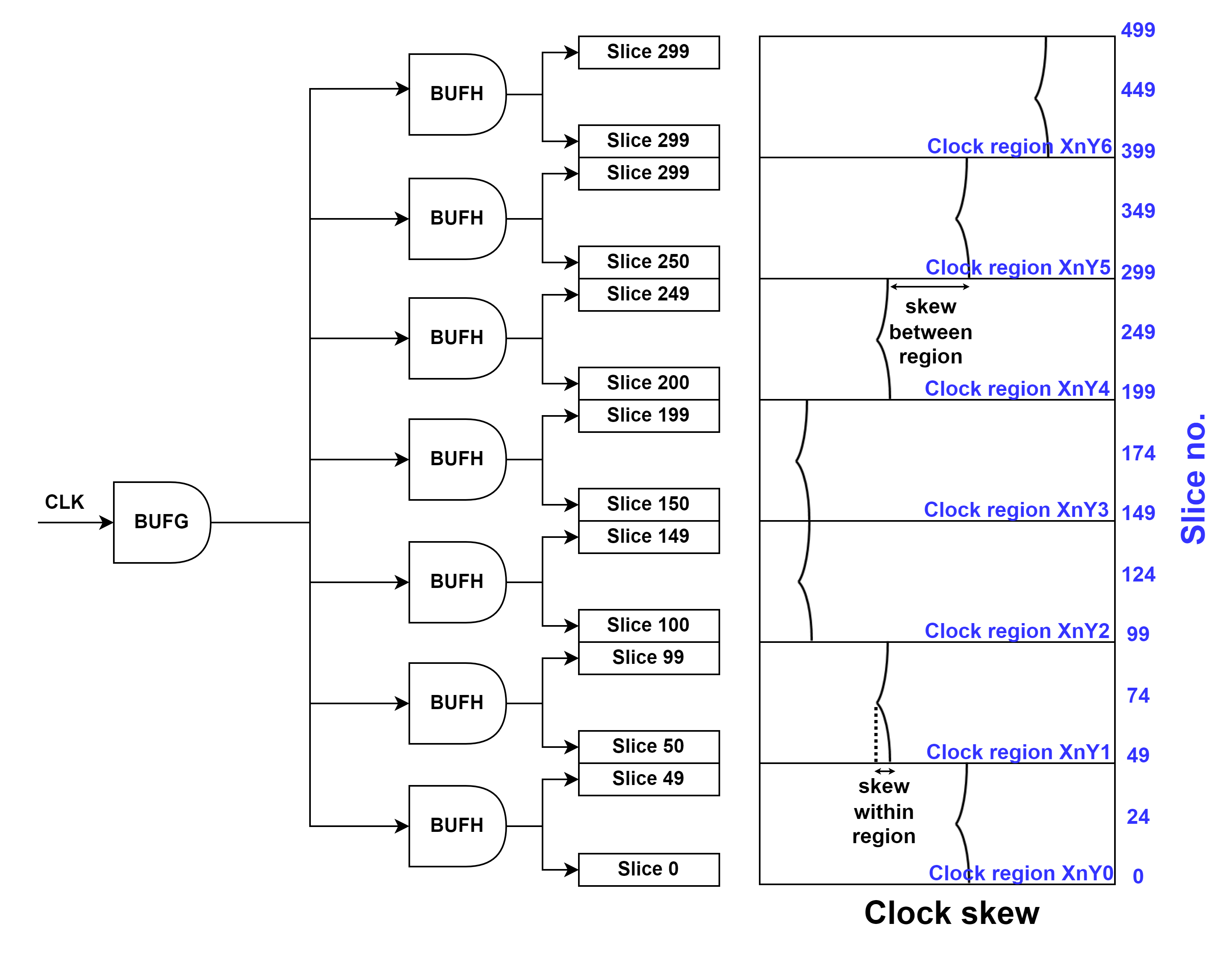}
\caption{\label{fig:4} The relationship between clock region and skew.}
\end{figure}

The relationship between clock region and skew of Xilinx kinex-7 series FPGAs is shown in figure \ref{fig:4}. For 400 bits tapped delay line, the clock region skew below the BUFG will not cause severe bubbles, but will shrink the pulse. The clock region skew above the BUFG will cause severe bubbles, which is determined by the order in which the clock arrives at two clock regions of the same delay chain. For the de-bubble scheme, only the delay chain above the BUFG needs to be processed with severe bubble.

The clock skew at the edge of the clock region can reach 173 ps\cite{RN26}, which will lead to severe bubble with a length of more than 10. Fortunately, severe bubble is only generated when the boundary of two clock regions is crossed by delay taps. Therefore, the original bits code of the range near the clock region boundary can be checked to determine whether there is a severe bubble and carry out corresponding processing. This processing will be implemented before entirely encoding and will not affect the taps outside the range near the clock region boundary. For 400 bits tapped delay line, the range to be checked is near 200th tap.

Clock skew can cause bubbles\cite{RN25}. Severe bubbles are created when there is a significant skew in timing between clock regions and the tapped delay line crosses the boundary between two such regions. Severe bubbles only happen on both sides of the clock region boundary at the same time. This is because if bubbles were to occur within the same clock region, there would be no clock region skew to cause severe bubble.

\begin{figure}[htbp]
    \centering
    \begin{subfigure}{0.49\linewidth}
        \centering
        \includegraphics[width=\linewidth]{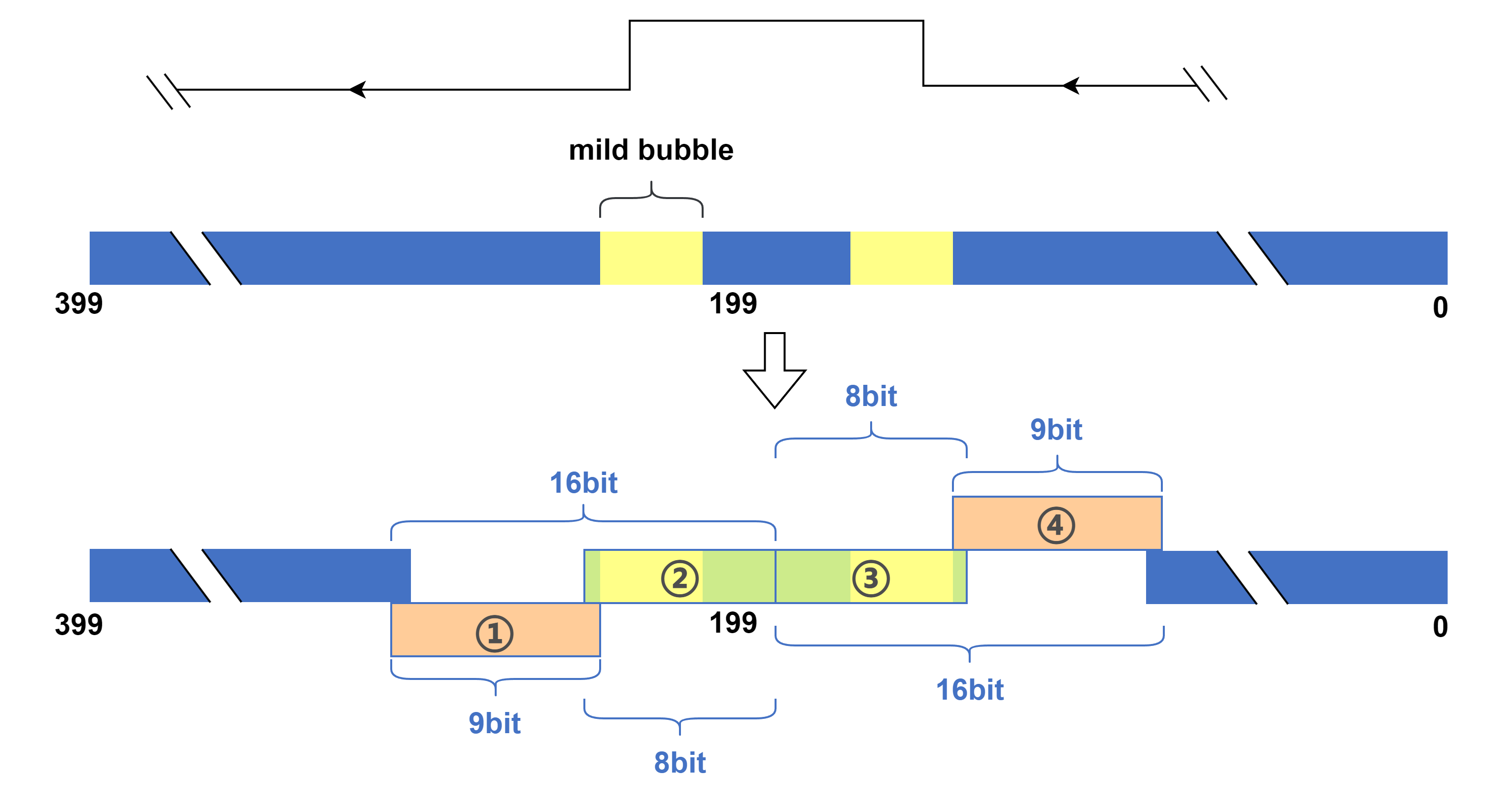}
        \caption{}
        \label{fig:6a}
    \end{subfigure}
    \begin{subfigure}{0.49\linewidth}
        \centering
        \includegraphics[width=\linewidth]{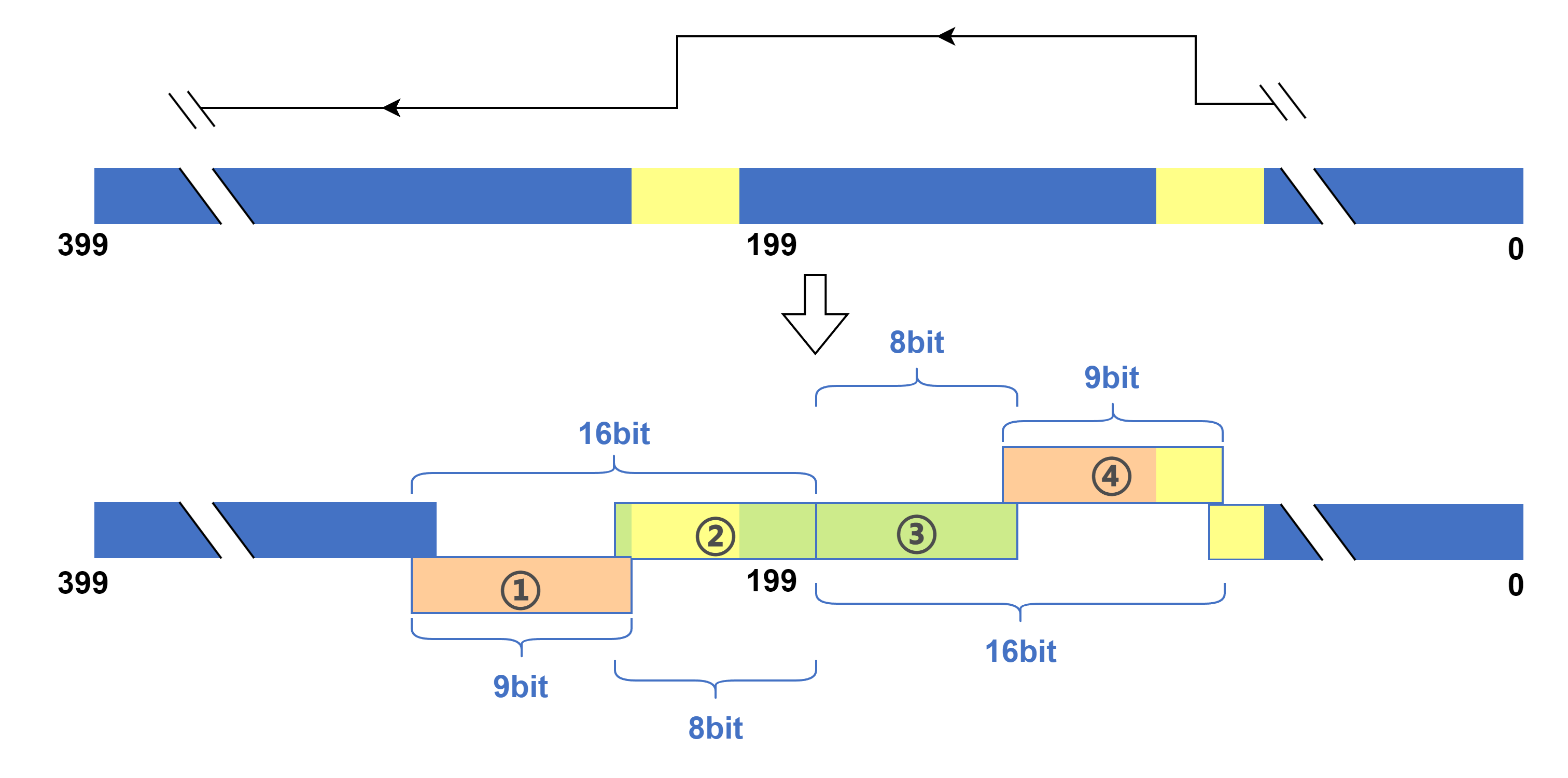}
        \caption{}
        \label{fig:6b}
    \end{subfigure}


    \begin{subfigure}{0.49\linewidth}
        \centering
        \includegraphics[width=\linewidth]{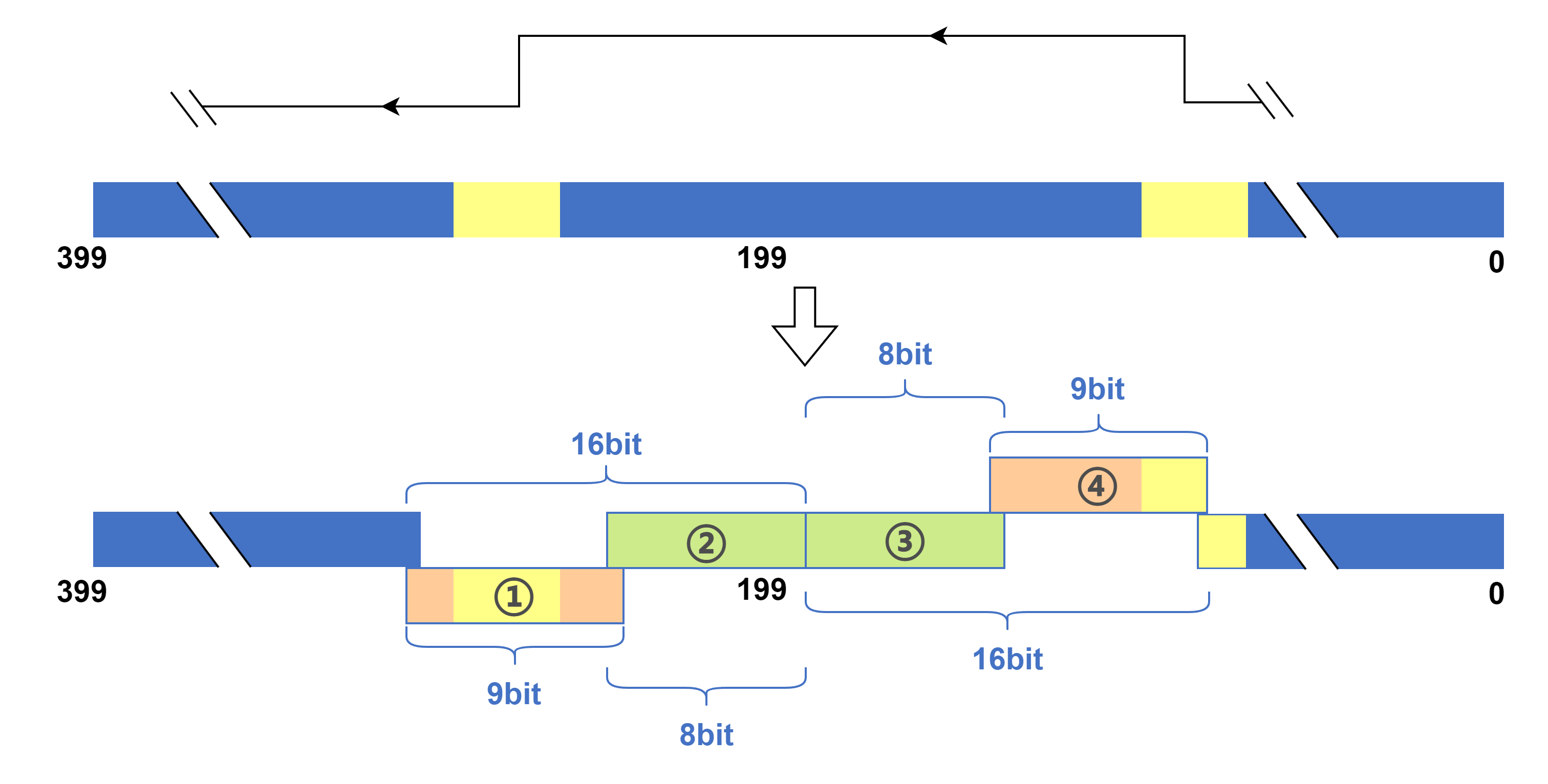}
        \caption{}
        \label{fig:6c}
    \end{subfigure}
    \begin{subfigure}{0.49\linewidth}
        \centering
        \includegraphics[width=\linewidth]{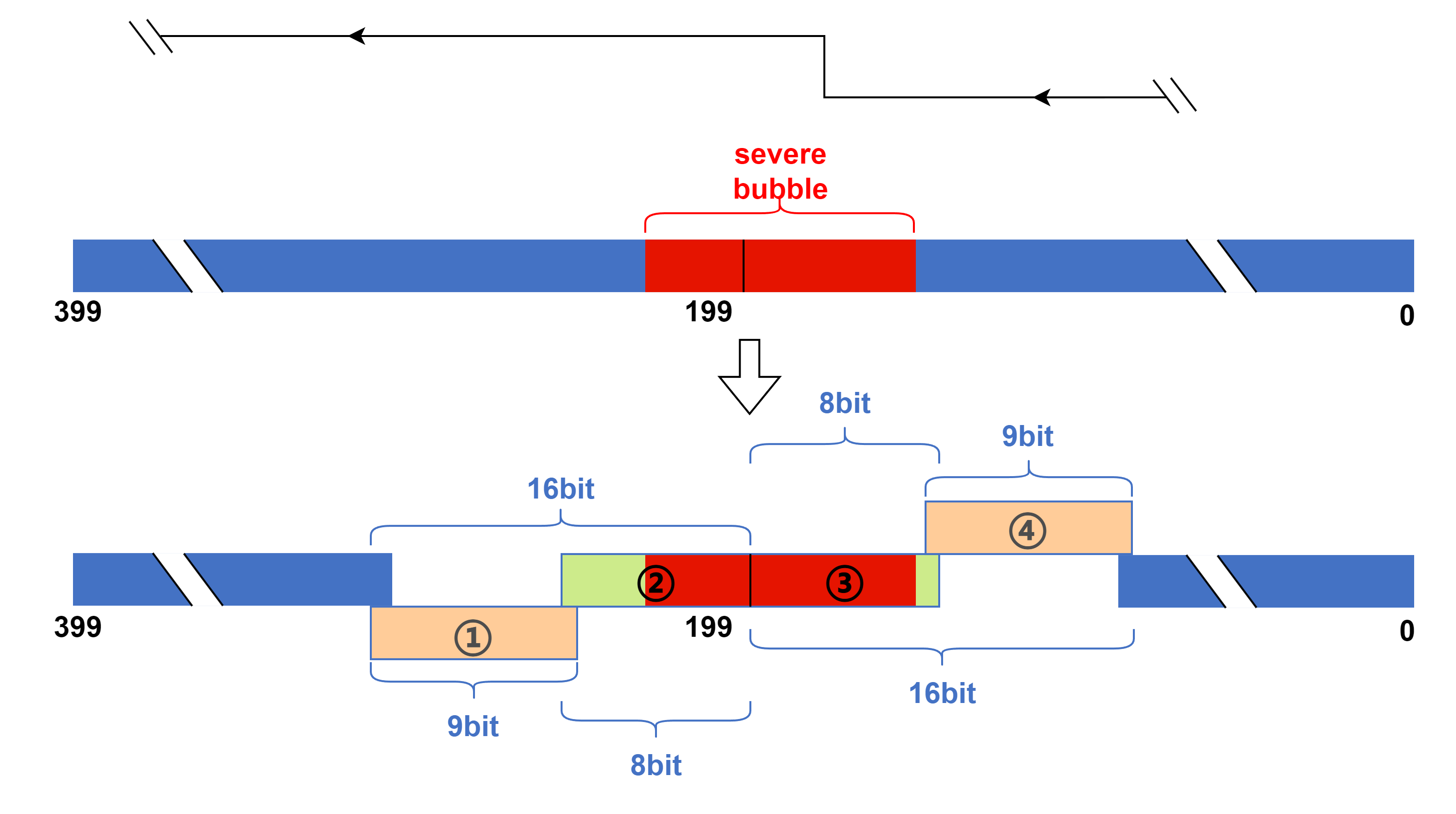}
        \caption{}
        \label{fig:6d}
    \end{subfigure}


    \begin{subfigure}{0.45\linewidth}
        \centering
        \includegraphics[width=\linewidth]{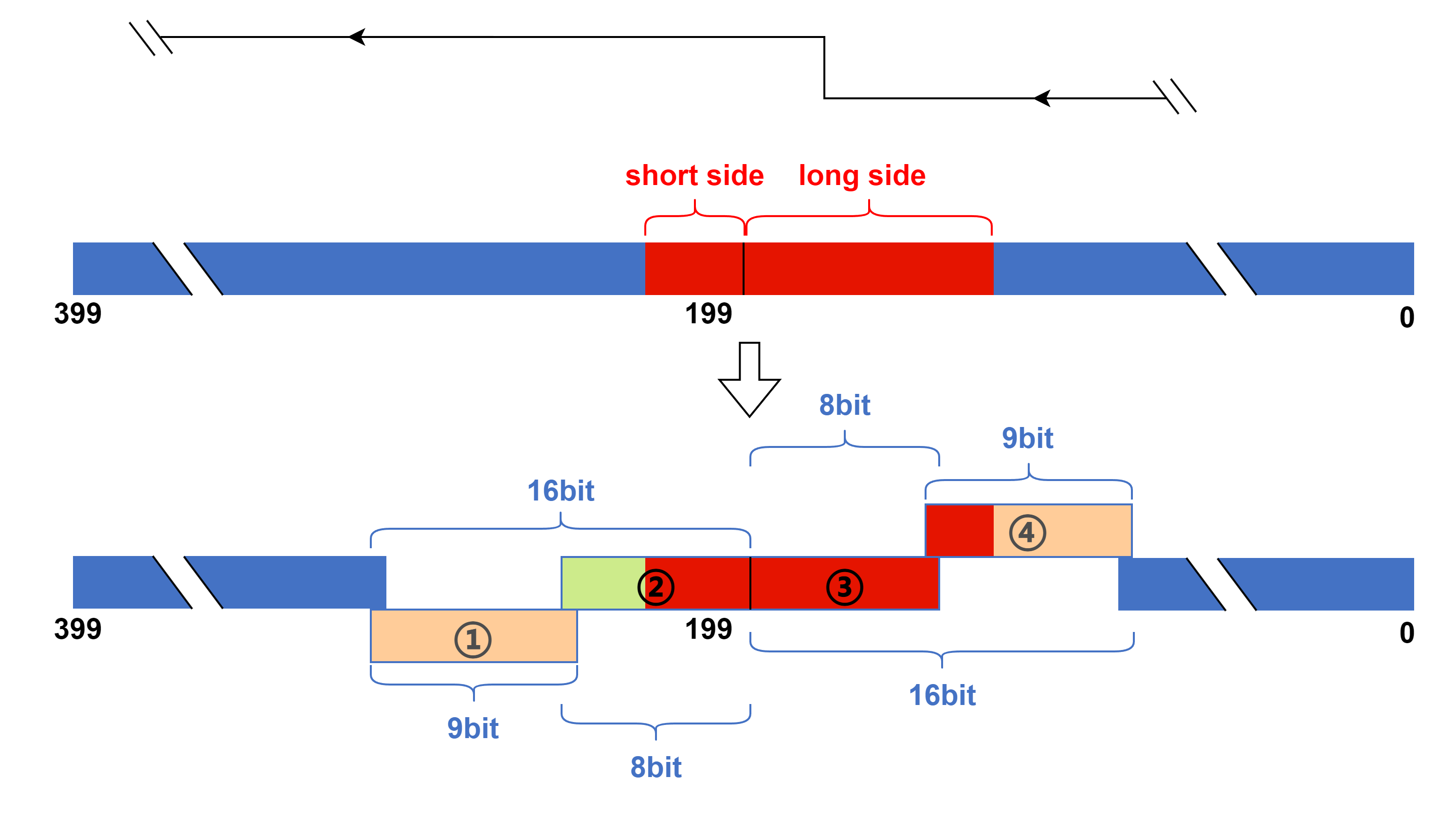}
        \caption{}
        \label{fig:6e}
    \end{subfigure}
    
\caption{Several cases when the edges appear near 200th bit. (a) Two general edges fall in region \textcircled{2}+\textcircled{3}. (b) Two general edges fall in region \textcircled{2}+\textcircled{4}. (c) Two general edges fall in region \textcircled{1}+\textcircled{4}. (d) Two general edges fall in region \textcircled{2}+\textcircled{3}. (e) Two general edges fall in region \textcircled{2}+\textcircled{3}+\textcircled{4}.}
\label{fig:6}
\end{figure}

For severe bubbles, we label the side closer to the 200th tap as the "short side" and the side farther from the 200th tap as the "long side," as shown in figure \ref{fig:6b}. If both sides are the same distance from the 200th tap point, they are both called "short sides." 

In our test, we have not observed any instances of a "short side" with a length exceeding 5 taps, a "long side" with a length exceeding 11 taps, or severe bubbles with a length exceeding 12 taps. To be cautious, we define the "short side" as being less than 8 taps, the "long side" as less than 15 taps, and severe bubbles as less than 16 taps. To avoid confusion between bubbles and the edges of the signal, the distance between edges should be ensured at a minimum of 16 taps. 

Based on the possible range of severe bubble, the taps near the 200th tap can be divided into four parts: the close 8-bit segments [207:200] and [199:192], and the distant 7-bit segments [214:208] and [191:185]. By checking for "1-0" or "0-1" transitions in these regions, the presence of a severe bubble can be ascertained. To recognize the transition edges at [215:214], [208:207], [192:191], and [185:184], the distant 7-bit segments are extended to 9-bit segments: [215:207] and [182:184], as shown in figure \ref{fig:6}.
\begin{figure}[htbp]
\centering 
\includegraphics[width=0.7\linewidth]{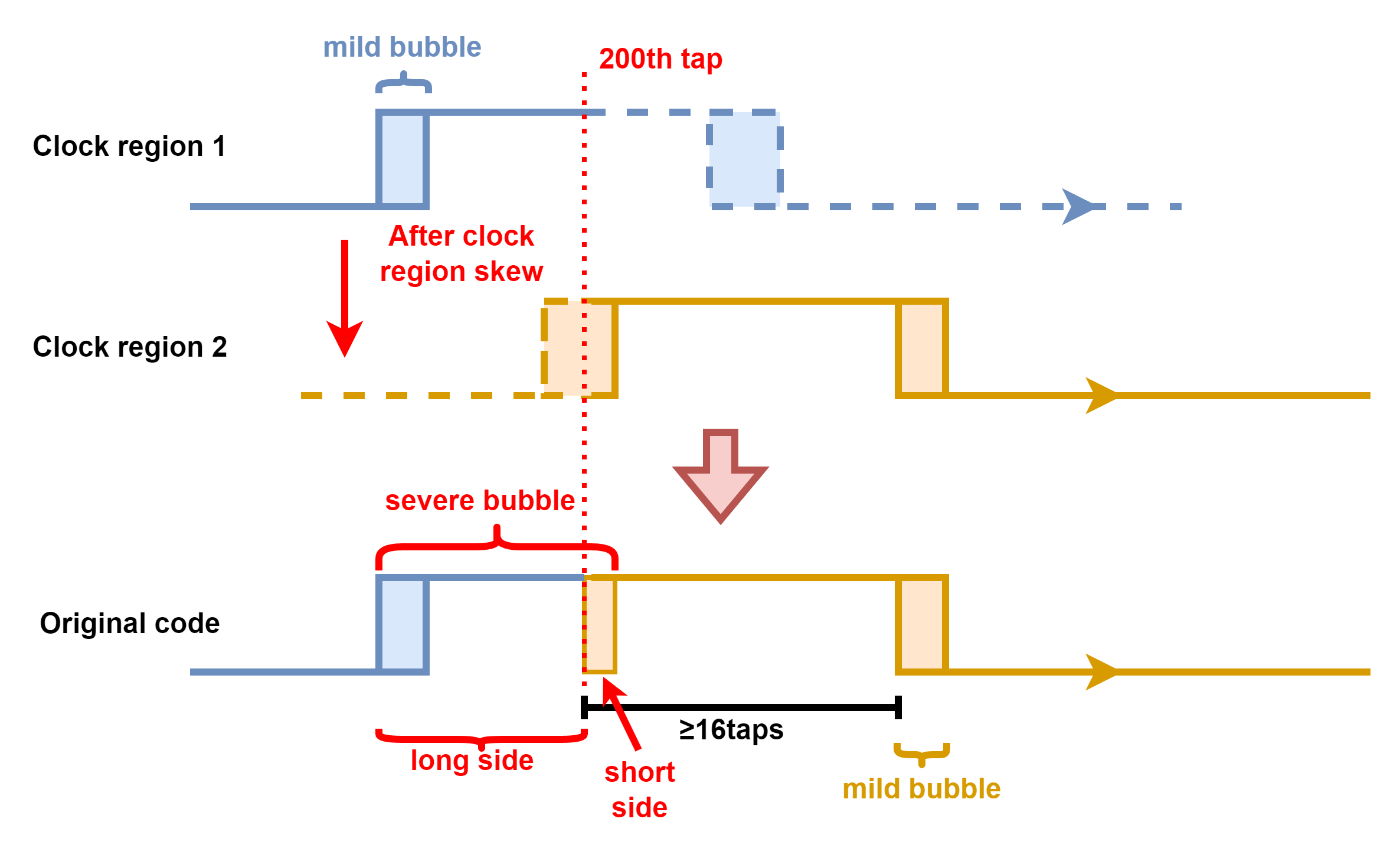}
\caption{\label{fig:5}  The process of the severe bubble generation.}
\end{figure}

The generation process of the severe bubble is shown in figure \ref{fig:5}. Because of the region clock skew, the tapped delay line sampled by clock region 2 is delayed by more than 100 ps compared with clock region 1, so a severe bubble is generated in original code. In order to avoid confusion caused by two types of bubbles, sever and mile, falling within the inspection region [215:185] at the same time, the distance between edges must be large enough. Figure \ref{fig:5} shows an extreme case. The rising edge after clock region skew just falls near the 200th tap, forming a severe bubble with the largest long side and the shortest short side. At this time, in order to ensure that the normal falling edge does not fall within the region of 16 taps near the 200th tap, the distance between the two edges is required to be no less than 16+6+6=28 taps. Where 6 is the maximum depth of mile bubbles\cite{RN24}.

Figure \ref{fig:6} illustrates a recognition scheme for severe bubble. When different edges states approach the vicinity of the 200th tap, the states of the four regions also vary. For ease of explanation, four areas to be inspected are labelled as \textcircled{1}, \textcircled{2}, \textcircled{3}, and \textcircled{4}, as shown in figure \ref{fig:6}. When transition such as "0-1" or "1-0" is detected in a region (e.g., region \textcircled{1}), the region is considered to be in a "true" state.

When \textcircled{1}\&\textcircled{4} = "true", corresponding to figure \ref{fig:6c}, the transitions have occurred outside the range where a severe bubble is likely to happen, indicating the presence of multiple edges. If the distance between edges is no less than 28 taps, a severe bubble does not exist.

Else, when \textcircled{2}\&\textcircled{4} = "true", corresponding to image figure \ref{fig:6b} or figure \ref{fig:6e}, this suggests the possibility of a severe bubble, but it could also be mild bubbles caused by two ordinary edges. The presence of severe bubble can be determined by checking the values at bits 207 and 199-16-6+1=178. If they are different, a severe bubble exists, if they are the same, there are two ordinary edges. (207 represents the adjacent bit between regions \textcircled{1} and \textcircled{2}, and 6 is the maximum depth of mild bubbles).

Else, when \textcircled{1}\&\textcircled{3} = "true", the processing is akin to when \textcircled{2}\&\textcircled{4} = "true". The bits to compare are 200+16+6-1=221 and 192. 

Else, when both \textcircled{2}\&\textcircled{3} = "true", corresponding to image figure \ref{fig:6a} or figure \ref{fig:6d}, if the distance between edges is no less than 28 taps, a severe bubble can be confirmed to exist.

\begin{figure}[htbp]

    \centering
    \begin{subfigure}{\linewidth}
        \centering
        \includegraphics[width=0.7\textwidth]{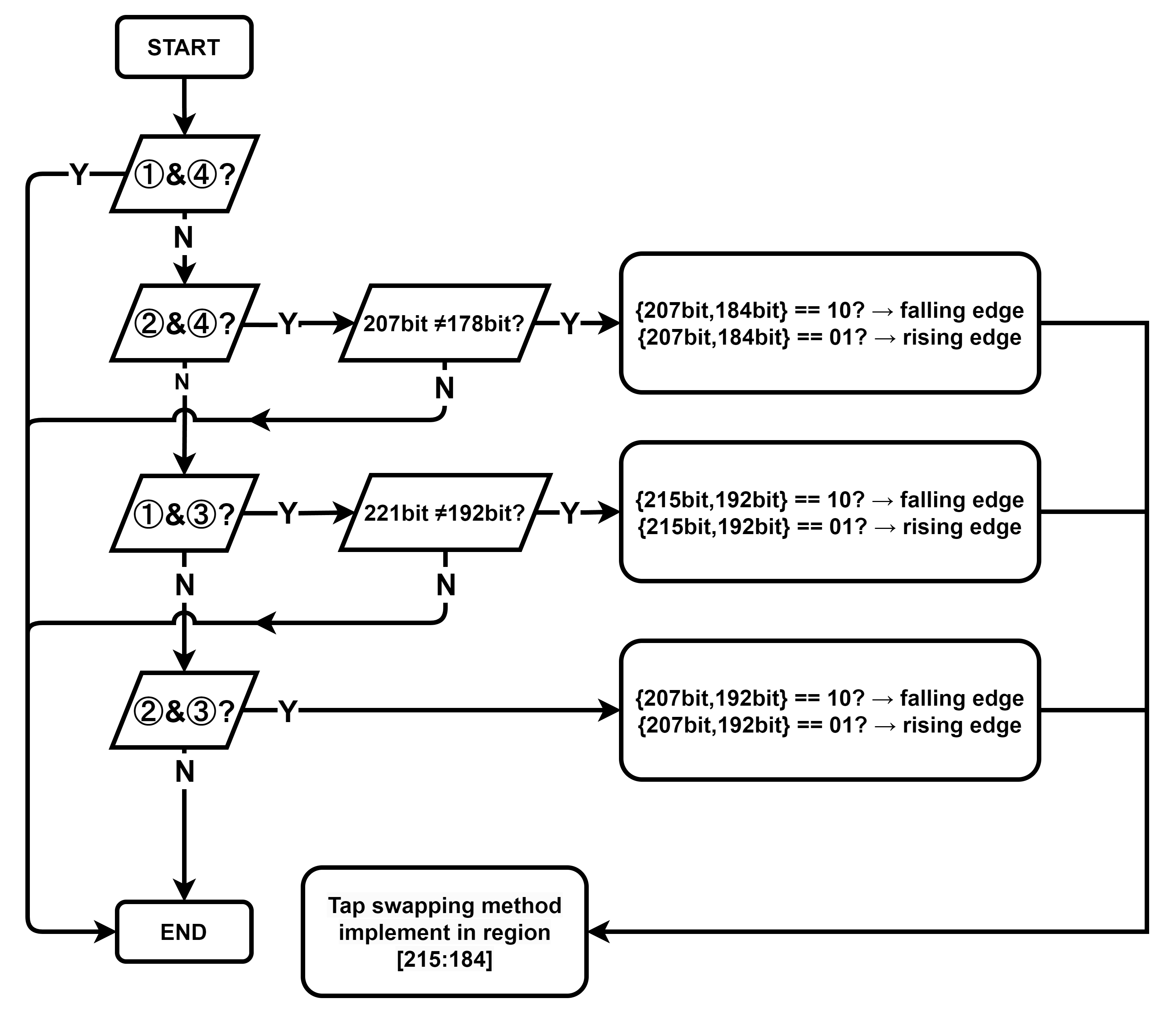}
        \caption{}
        \label{fig:7a}
    \end{subfigure}
    
    \begin{subfigure}{\linewidth}
        \centering
        \includegraphics[width=0.7\textwidth]{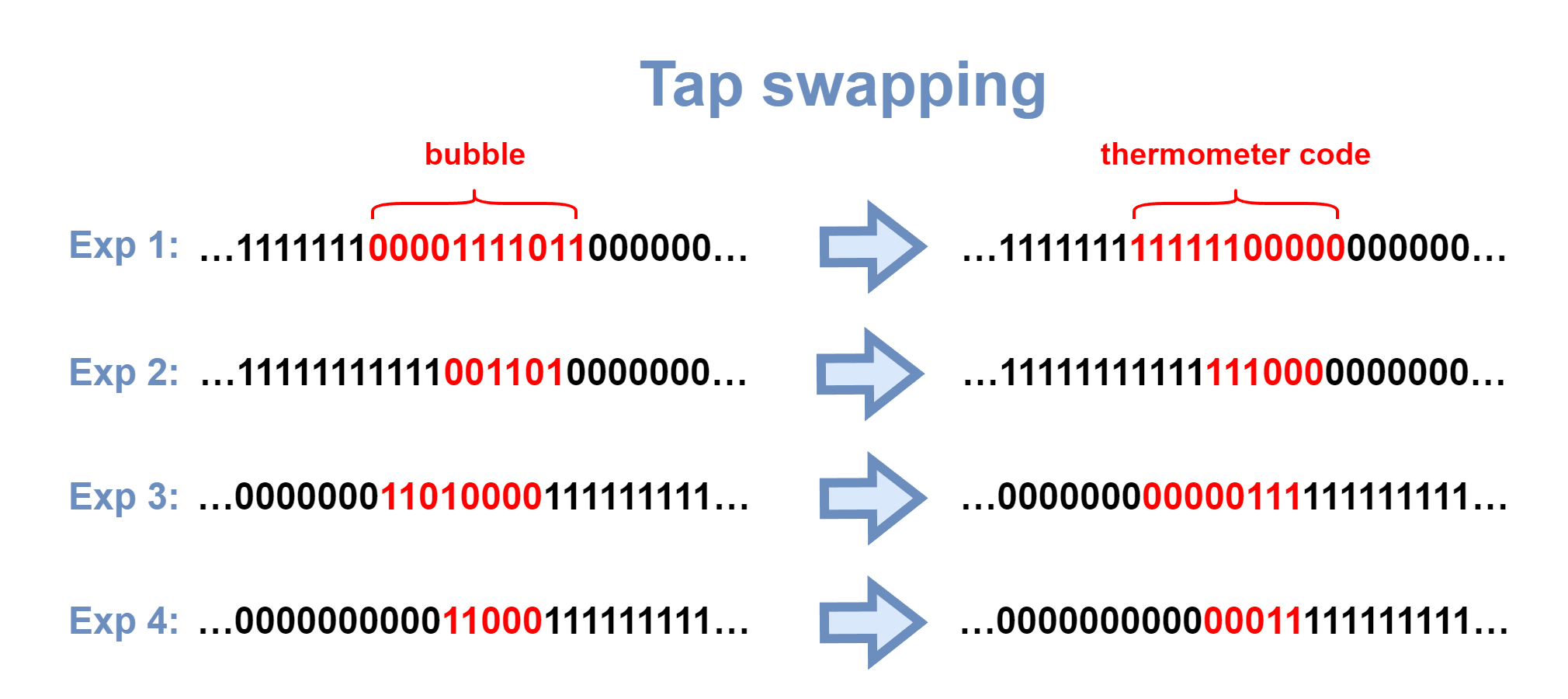}
        \caption{}
        \label{fig:7b}
    \end{subfigure}
    
\caption{(a) The process of severe bubble solution. (b) Several examples of the tap swapping method.}
\label{fig:7}
\end{figure}

The complete flowchart for the severe bubble solution is shown in figure \ref{fig:7a}. When the spacing between edges exceeds 28 taps, there will not be both mild and severe bubble edges present in the detection region simultaneously, which could cause interference. Therefore, when confirming the existence of a severe bubble, a tap swapping method can be directly implemented in the region [215:184], where the severe bubble occurs. Due to the identical encoding results between tap swapping and the ones-counter encoder \cite{RN22}, implementing the tap swapping method simply requires counting the number of '1's or '0's and converting the original taps to thermometer code online, which is shown in the example figure \ref{fig:7b}. The implementation of the tap swapping method needs to determine whether the edge is a rising or falling edge, which can be obtained at the boundary of the region where severe bubble is detected.

Due to the inability to perform multi-edge decomposition encoding simultaneously during the processing of severe bubbles, additional dead time may be introduced. One could consider implementing the severe bubble solution as a pipeline to eliminate dead time. However, this approach would require buffering 400 taps per clock cycle, which consumes a lot of resources. Fortunately, although the logic of the severe bubble solution is intricate, it only operates on 32 taps, allowing for rapid completion without causing timing violations. In this work, severe bubble solution does not utilize pipelining, with a completion time of 2 clock cycles, which is 2*5 = 10 ns, exceeding the dead time of stretcher module and introducing additional dead time.

As depicted in figure \ref{fig:3}, after the resolution of severe bubbles, the tapped delay line is left with only shorter mild bubbles. The multi-edge decomposition encoder can pipeline the processing and encoding of the code with mild bubbles without incurring additional dead time. The pipeline requires two clock cycles, totaling a 2+2=4 encoding clock cycle.

\subsection{Offline calibration module}
The calibration module is set to be offline, which is due to the current application scenario where the trigger rate of sampling signal is not high. However, the system still retains the capability to implement online calibration module. The calibration scheme is universal for TDL-TDC and is named the Code Density Test \cite{RN27}. Given FPGA TDL, composed of CARRY4 cascades, exhibits significant nonlinearity. Calculating fine time using the average LSB in bins would result in a considerable measurement error. Therefore, calibrating the width of each bin is more suitable for this type of TDL-TDC.

The principle of Code density test is to input a large number of random signals into TDL to generate a large number of encoding outputs, and perform statistical calculations on these encoding results to calibrate the time width of each encoding value. Because the sampling clock and signal are asynchronous, when the sampling clock arrives, the position of the signal in the TDL is random, so the delay time distribution of the signal in the TDL is uniform. The number of random signals hitting a bin is directly proportional to its bin width (for this work, ‘bin’ here represents a result of the encoder, not a physical delay cell), In this way, the time measurement of bin width can be transformed into the measurement of the number of random signals falling into bins. When the number of random signals is large enough, it can be considered that the statistical value is the same as the actual value.

In this work, the software LabVIEW is employed for offline calibration and timestamp output, with data being transmitted from FPGA to PC via TCP protocol. 

Each TDC channel is initially tested with 100,000 random signals to ensure that the hits are evenly distributed across the TDL, which, upon completion, generates a calibration table that accounts for the width of each bin. When the next sampling signal arrives, the fine time can be generated by referring to the time on the calibration table based on the bin's position, and combined with the coarse time to form a timestamp. The calibration module is executed once each time the TDC powers up.

\section{TEST}
\subsection{Test of the module of severe bubble solution}
A 2-channel TDC is implemented below the BUFG, which is based on the Xilinx Kintex-7 series xc7k325tffg676-2, and does not use a severe bubble solution. Another 2-channel TDC is implemented above the BUFG, which is based on the same type of FPGA, and use a module of severe bubble solution.

\begin{figure}[htbp]
    \centering
    \begin{subfigure}{0.43\linewidth}
        \includegraphics[width=\linewidth]{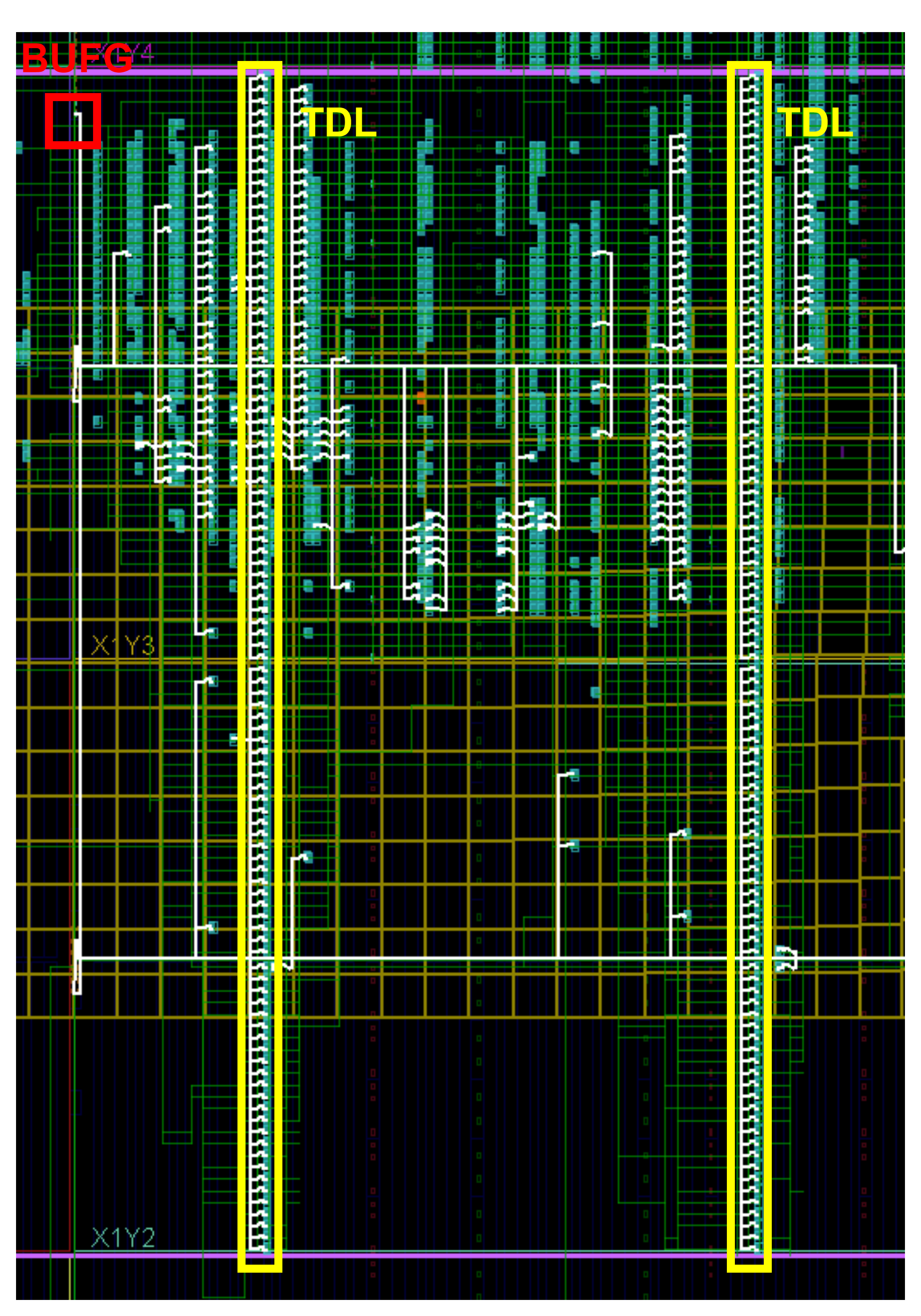}
        \caption{}
        \label{fig:8a}
    \end{subfigure}
    \begin{subfigure}{0.43\linewidth}
        \includegraphics[width=\linewidth]{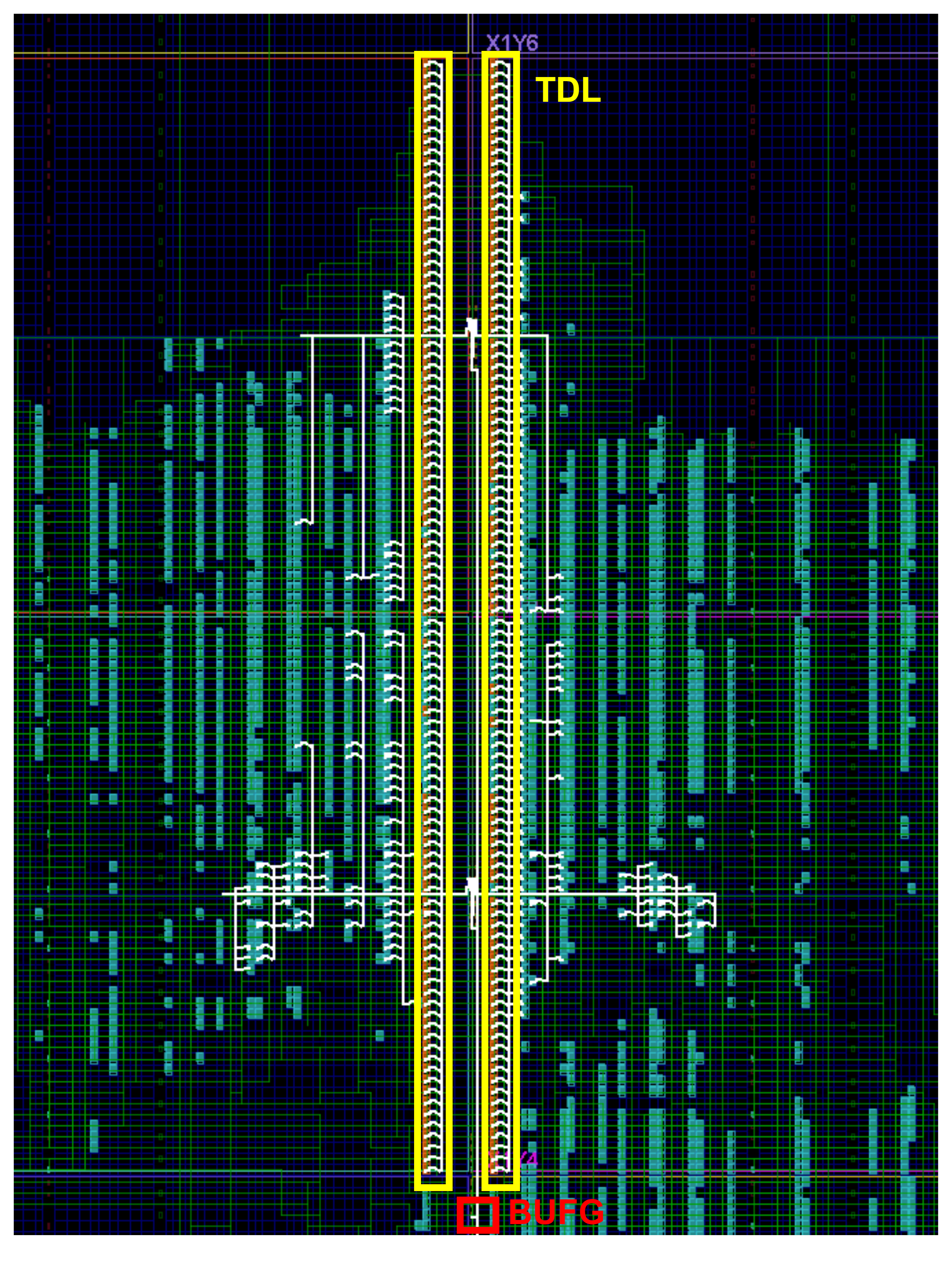}
        \caption{}
        \label{fig:8b}
    \end{subfigure}
    \caption{2-channel TDCs implemented in Xilinx Kintex-7 device. (a) 2-channel TDC below the BUFG without severe bubble solution module. (b) 2-channel TDC above the BUFG with severe bubble solution module.}
    \label{fig:8}
\end{figure}

The FPGA implementation of the TDC is shown in figure \ref{fig:8}, where the white part represents the sampling clock path, the red box encloses the used BUFG, the yellow box encloses the TDL along with the corresponding D flip-flops and clock paths, and outside the yellow box are the encoders, coarse counters, and other structures.

The FPGA original clock is provided by an external clock board (Si5345), which has a frequency of 100 MHz and is generated into 400 MHz sampling clock and 200MHz encoding clock through the internal PLL of FPGA. Linear power supply (GPC-3030DQ) voltage 12 V. 

The sampling signal used for testing the TDCs is generated by signal generator (AFG3252C) with a frequency of 500 Hz, high level 2.5 V, low level 0 V, and a duty cycle of 50\%. Sampling signal input from SMA port through 1 m coaxial cable, and the FPGA receiving level is LVCMOS25. 

\begin{figure}[htbp]
    \centering
    \begin{subfigure}{0.49\linewidth}
        \includegraphics[width=\linewidth]{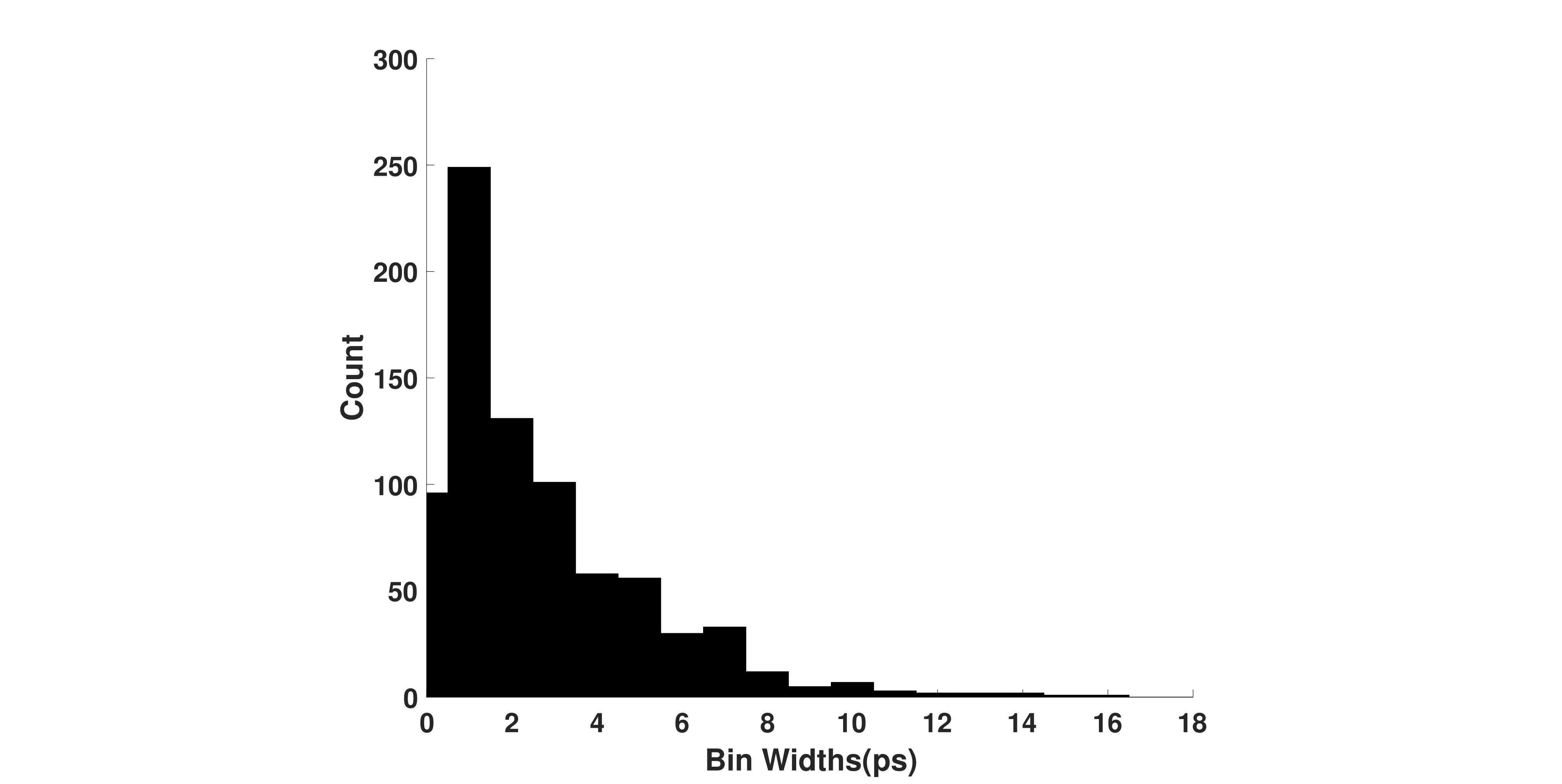}
        \caption{}
        \label{fig:9a}
    \end{subfigure}
    \begin{subfigure}{0.49\linewidth}
        \includegraphics[width=\linewidth]{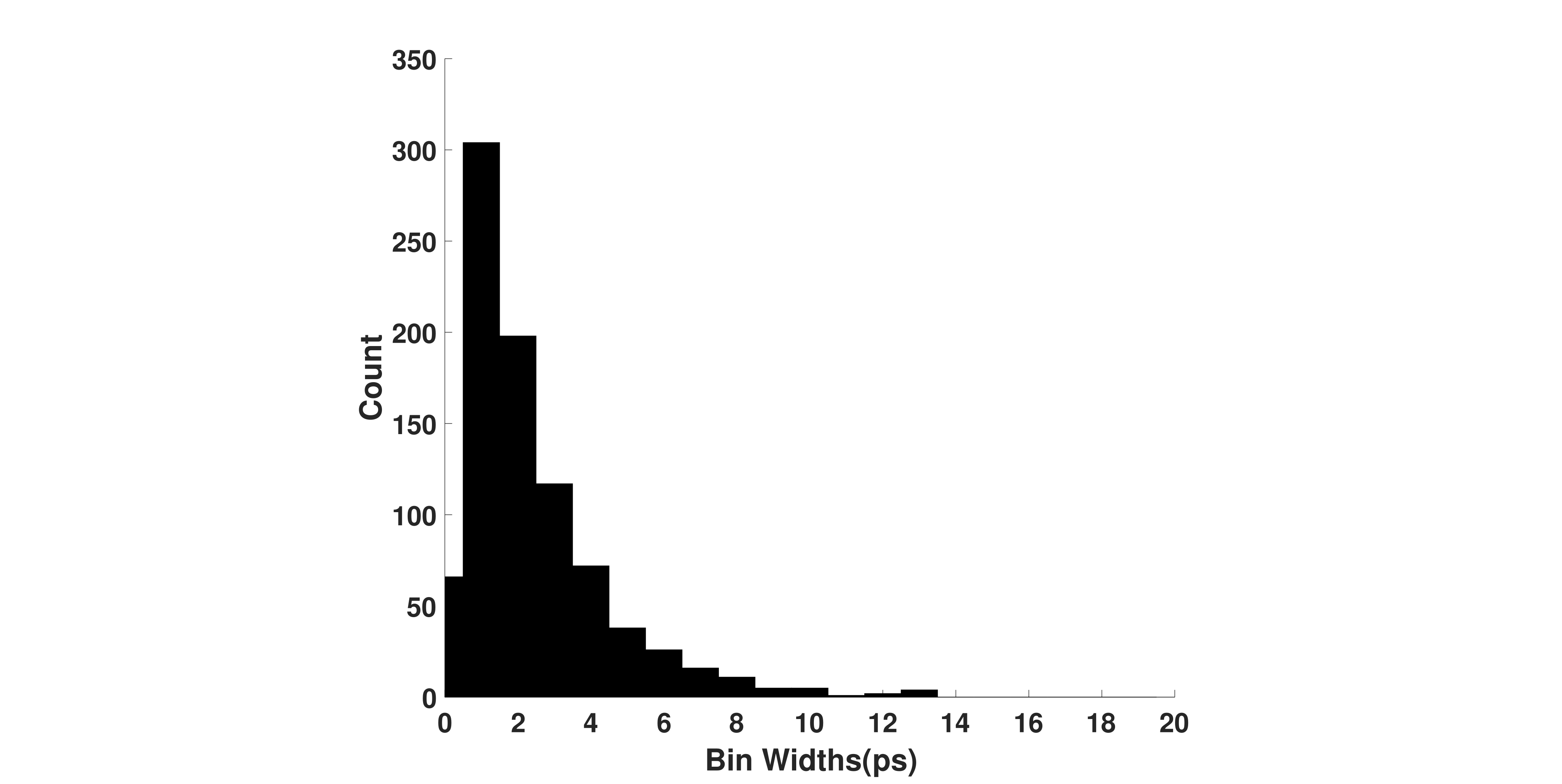}
        \caption{}
        \label{fig:9b}
    \end{subfigure}
    
    \begin{subfigure}{0.49\linewidth}
        \includegraphics[width=\linewidth]{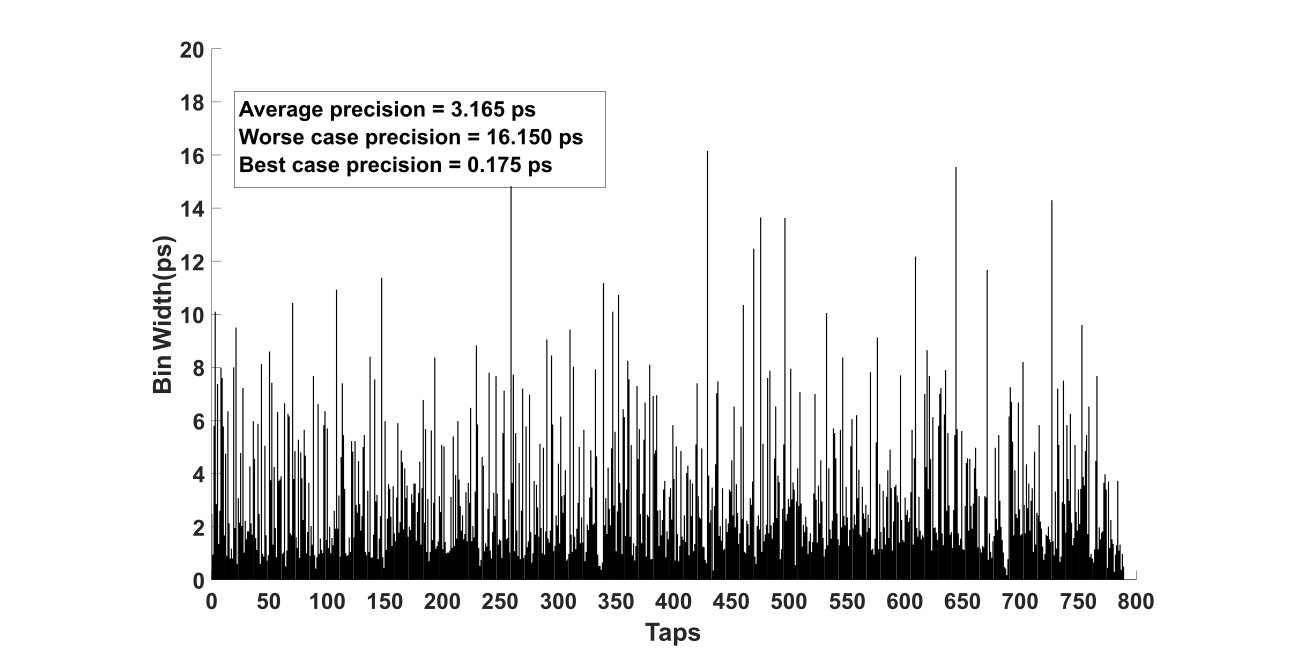}
        \caption{}
        \label{fig:9c}
    \end{subfigure}
    \begin{subfigure}{0.49\linewidth}
        \includegraphics[width=\linewidth]{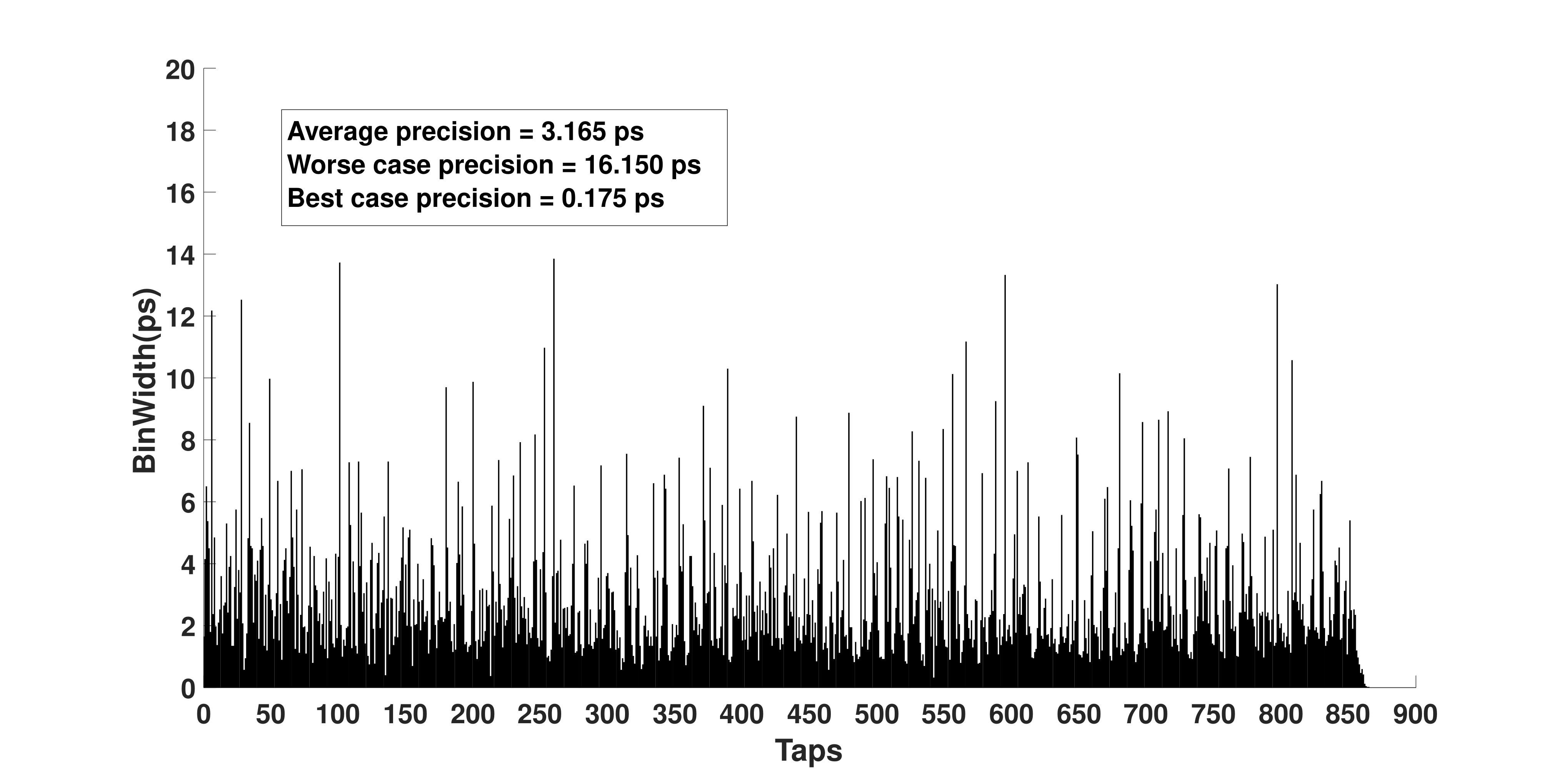}
        \caption{}
        \label{fig:9d}
    \end{subfigure}
    
    \begin{subfigure}{0.49\linewidth}
        \includegraphics[width=\linewidth]{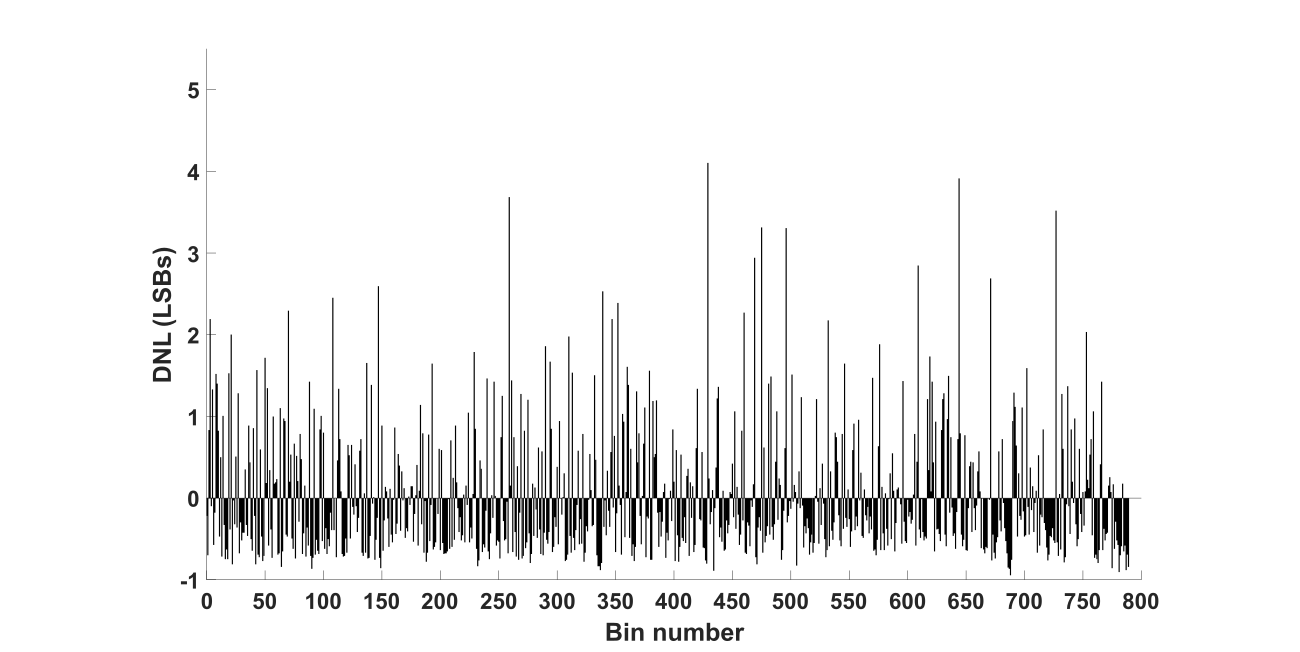}
        \caption{}
        \label{fig:9e}
    \end{subfigure}
    \begin{subfigure}{0.49\linewidth}
        \includegraphics[width=\linewidth]{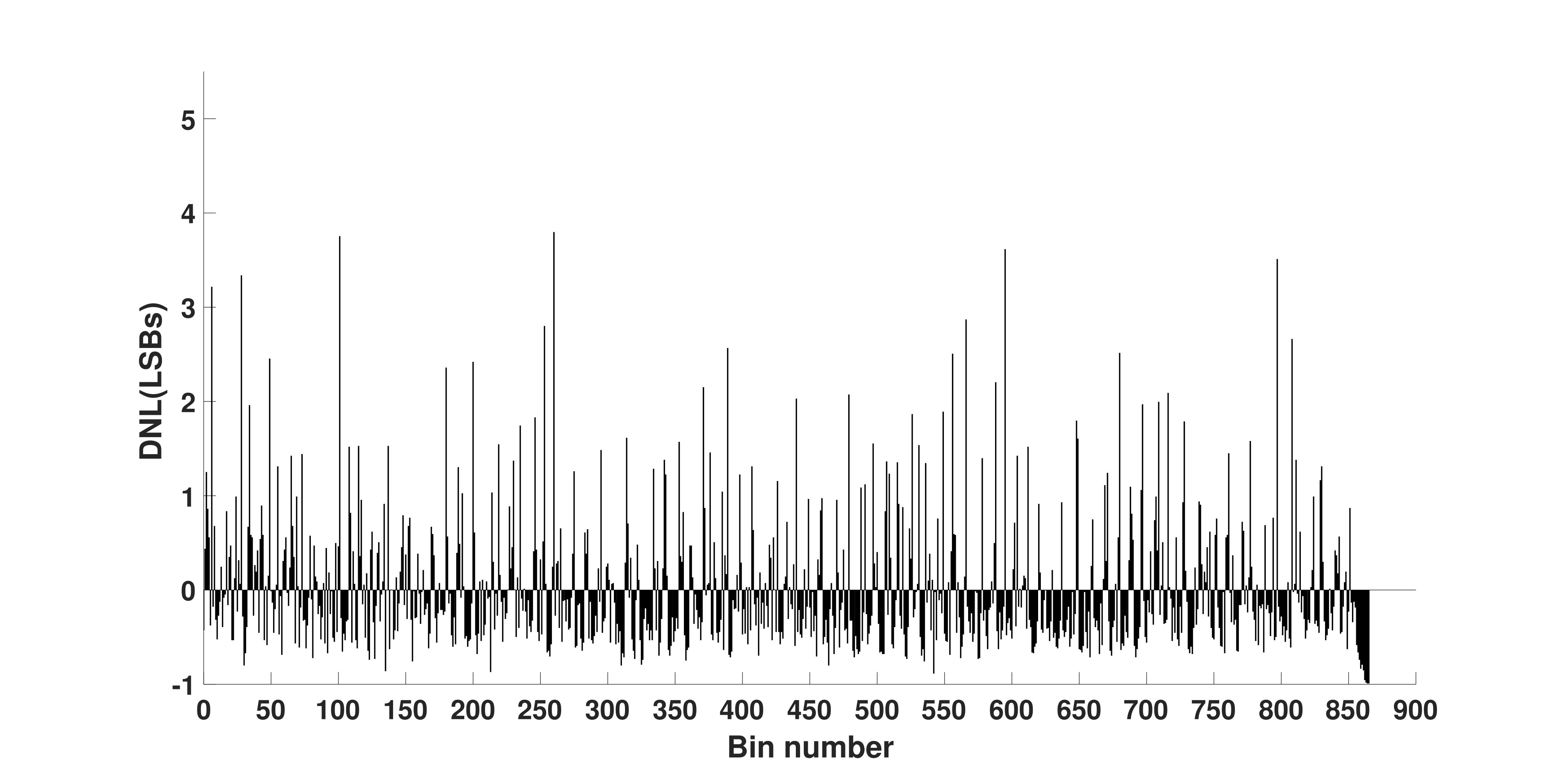}
        \caption{}
        \label{fig:9f}
    \end{subfigure}
    
    \begin{subfigure}{0.49\linewidth}
        \includegraphics[width=\linewidth]{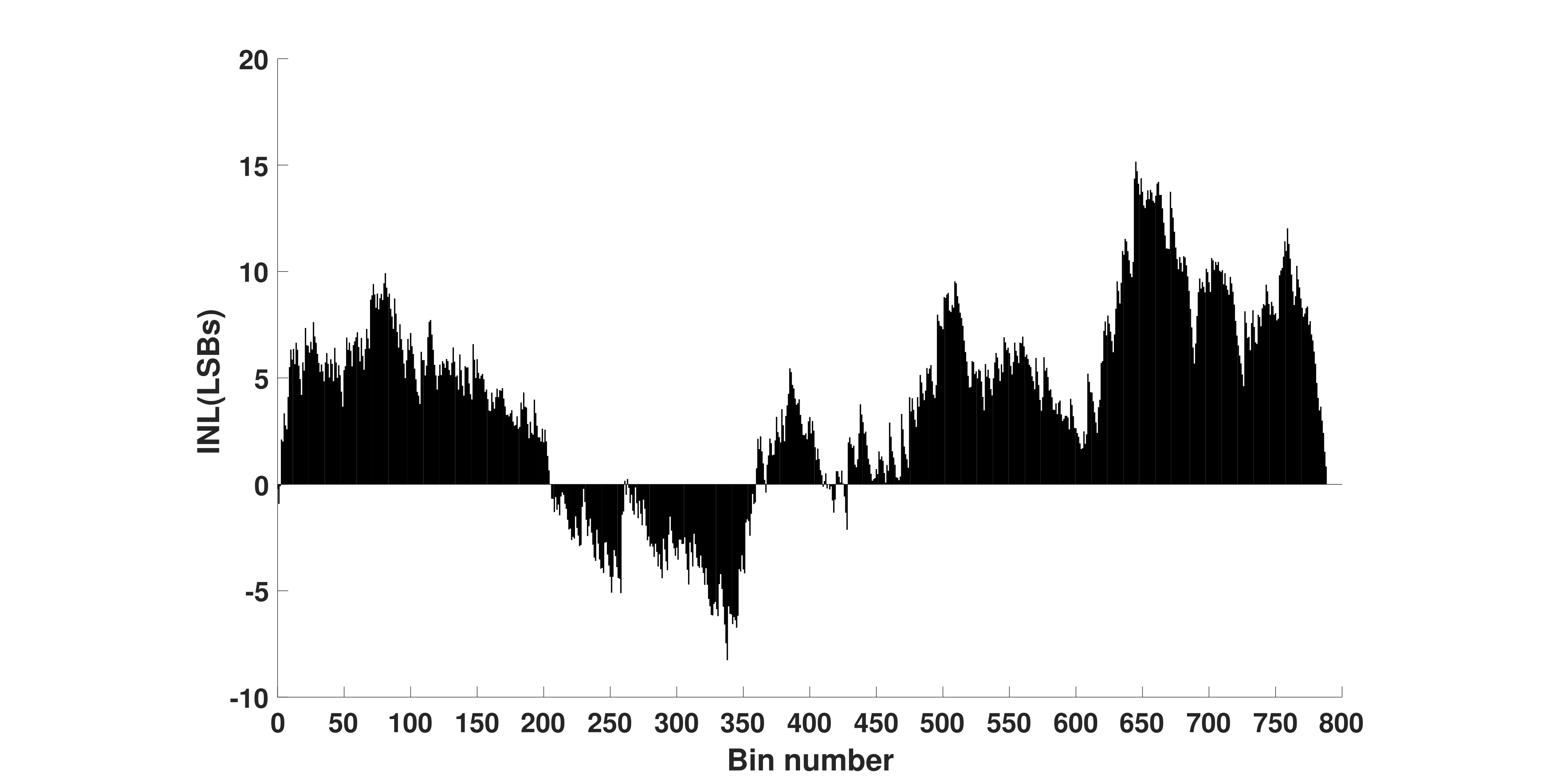}
        \caption{}
        \label{fig:9g}
    \end{subfigure}
    \begin{subfigure}{0.49\linewidth}
        \includegraphics[width=\linewidth]{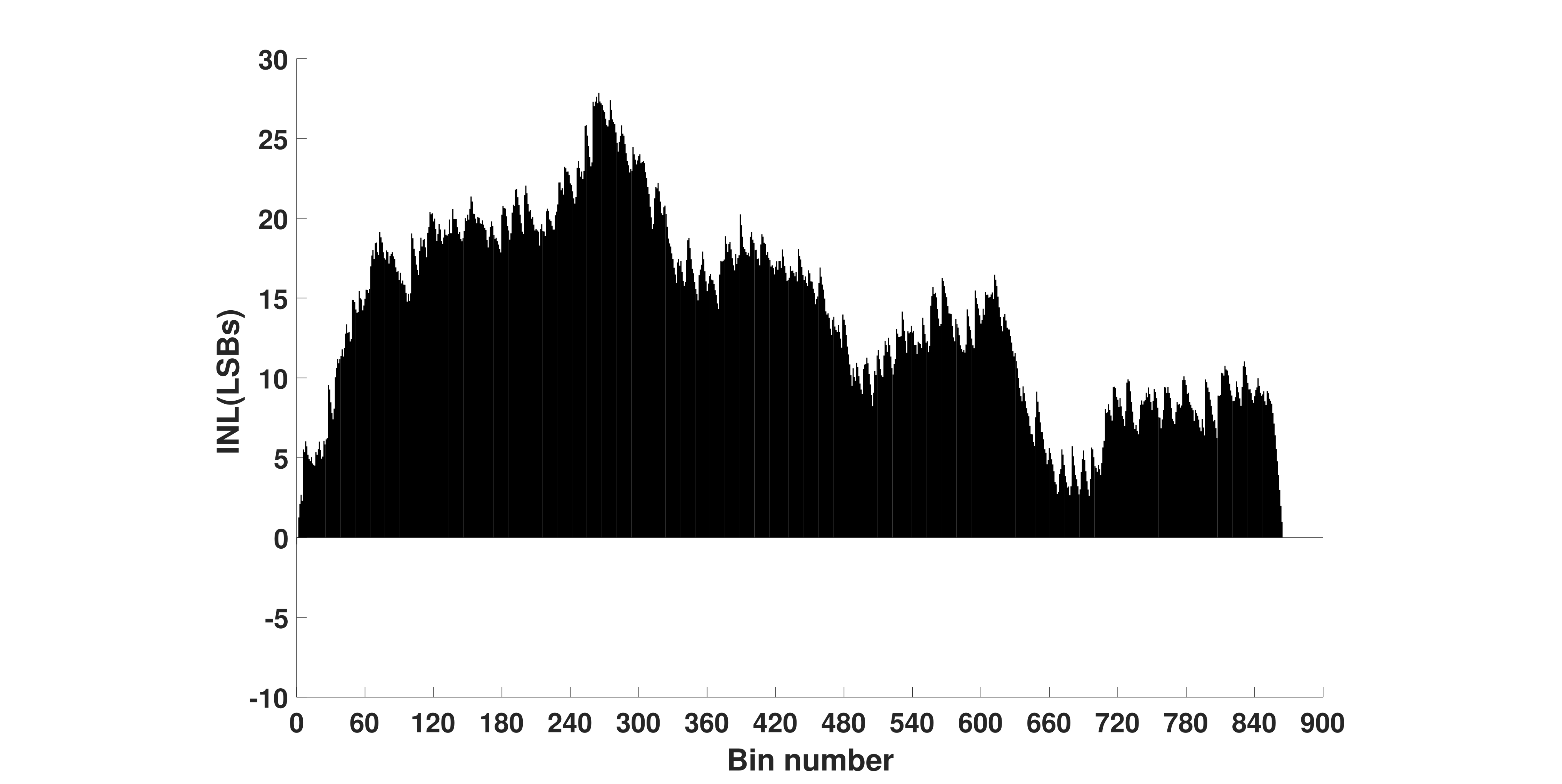}
        \caption{}
        \label{fig:9h}
    \end{subfigure}
    \caption{Test result of the 2-channel TDCs. (a) Count distribution of widths for channel \#0 of the TDC below BUFG. (b) Count distribution of widths for channel \#0 of the TDC above BUFG. (c) Taps width distribution for channel \#0 of the TDC located below BUFG. (d) Taps width distribution for channel \#0 of the TDC located above BUFG. (e) DNL for channel \#0 of the TDC located below BUFG. (f) DNL for channel \#0 of the TDC located above BUFG. (g) INL for channel \#0 of the TDC located below BUFG. (h) INL for channel \#0 of the TDC located above BUFG.}
    \label{fig:9}
\end{figure}

The test results are shown in figure \ref{fig:9}. The average Least Significant Bit (LSB) and Differential Non-Linearity (DNL) are generated from the taps data provided by the calibration module implemented in LabVIEW software, while the Integral Nonlinearity (INL) is derived by integrating the DNL values.

Figure \ref{fig:9c} represents the taps width distribution of channel \#0 of the TDC located below the BUFG, which has 790 taps.  Figure \ref{fig:9d} represents the width distribution of channel \#0 of the TDC located above the BUFG, which has 866 taps. Figure \ref{fig:9a} and figure \ref{fig:9b} respectively illustrate the count distribution of the widths for the 790 and 866 taps, with average LSBs of 3.165 and 2.887 ps. Figure \ref{fig:9e} and figure \ref{fig:9g} show the DNL and INL of the TDC channel \#0 located below the BUFG. Figure \ref{fig:9f} and figure \ref{fig:9h} show the DNL and INL of the TDC channel \#0 located above the BUFG. The DNL can reach a maximum of 4.10 and 3.80 times the average LSB, while the INL can reach a maximum of 15.16 and 27.9 times the average LSB. The measurement errors present in the INL are significant, but they can be entirely mitigated by bin-by-bin calibration. The temporal precision of the TDC is only determined by the DNL of the TDC bins\cite{RN24}.
\begin{figure}[htbp]
    \centering
    \begin{subfigure}{0.49\linewidth}
        \includegraphics[width=\linewidth]{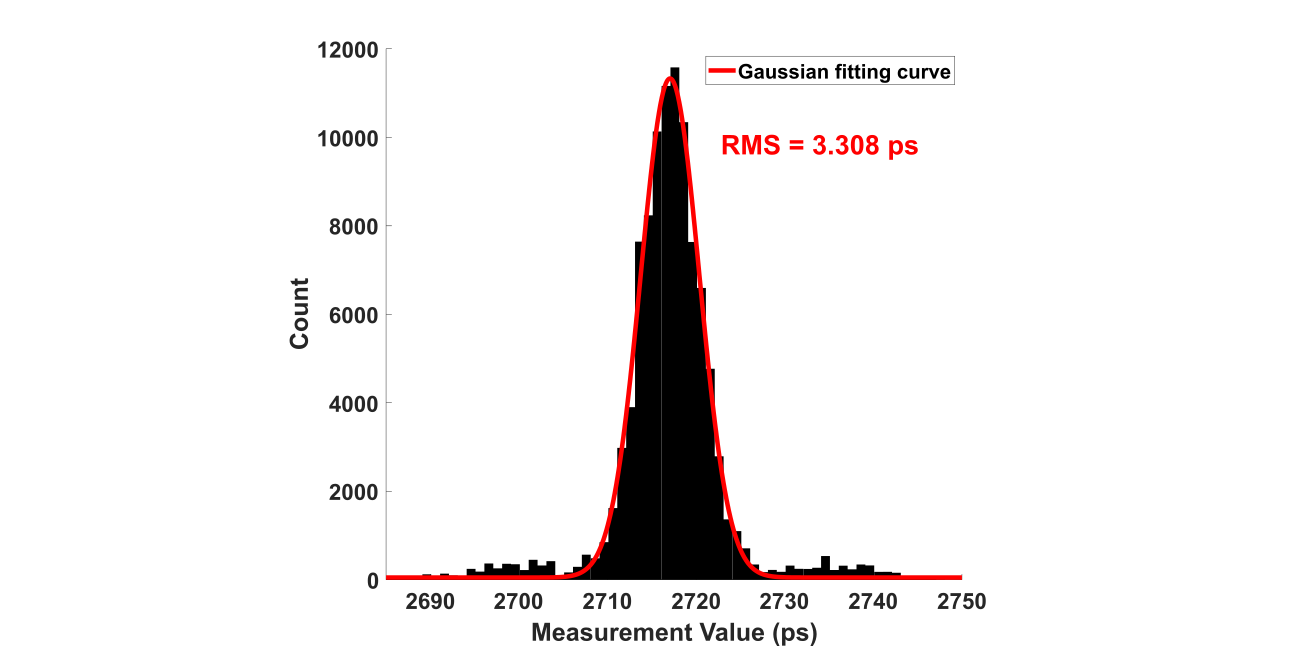}
        \caption{}
        \label{fig:10a}
    \end{subfigure}
    \begin{subfigure}{0.49\linewidth}
        \includegraphics[width=\linewidth]{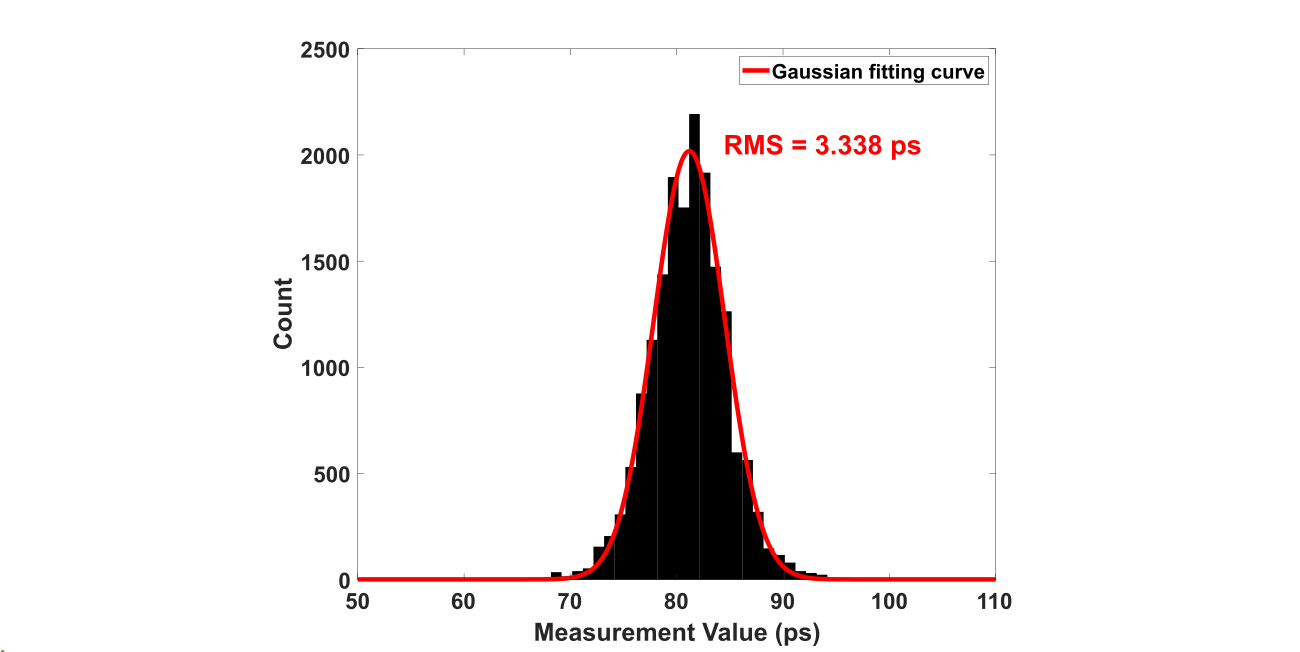}
        \caption{}
        \label{fig:10b}
    \end{subfigure}
    \caption{RMS of the 2-channel TDCs. (a) RMS of the 2-channel TDC located below BUFG. (b) RMS of the 2-channel TDC located above BUFG.}
    \label{fig:10}
\end{figure}

The coincidence time resolution testing result of the two types of 2-channel TDCs is shown in figure \ref{fig:10}. The 2-channel TDC located below BUFG had a root mean square (RMS) of 3.305 ps in each single channel, with each channel recording 103,744 sampling hits. The 2-channel TDC located above BUFG had an RMS of 3.338 ps in each single channel, with each channel recording 17,253 sampling hits. And the FPGA temperature was 46.6 °C.

\begin{figure}[htbp]
    \centering
    \includegraphics[width=0.4\linewidth]{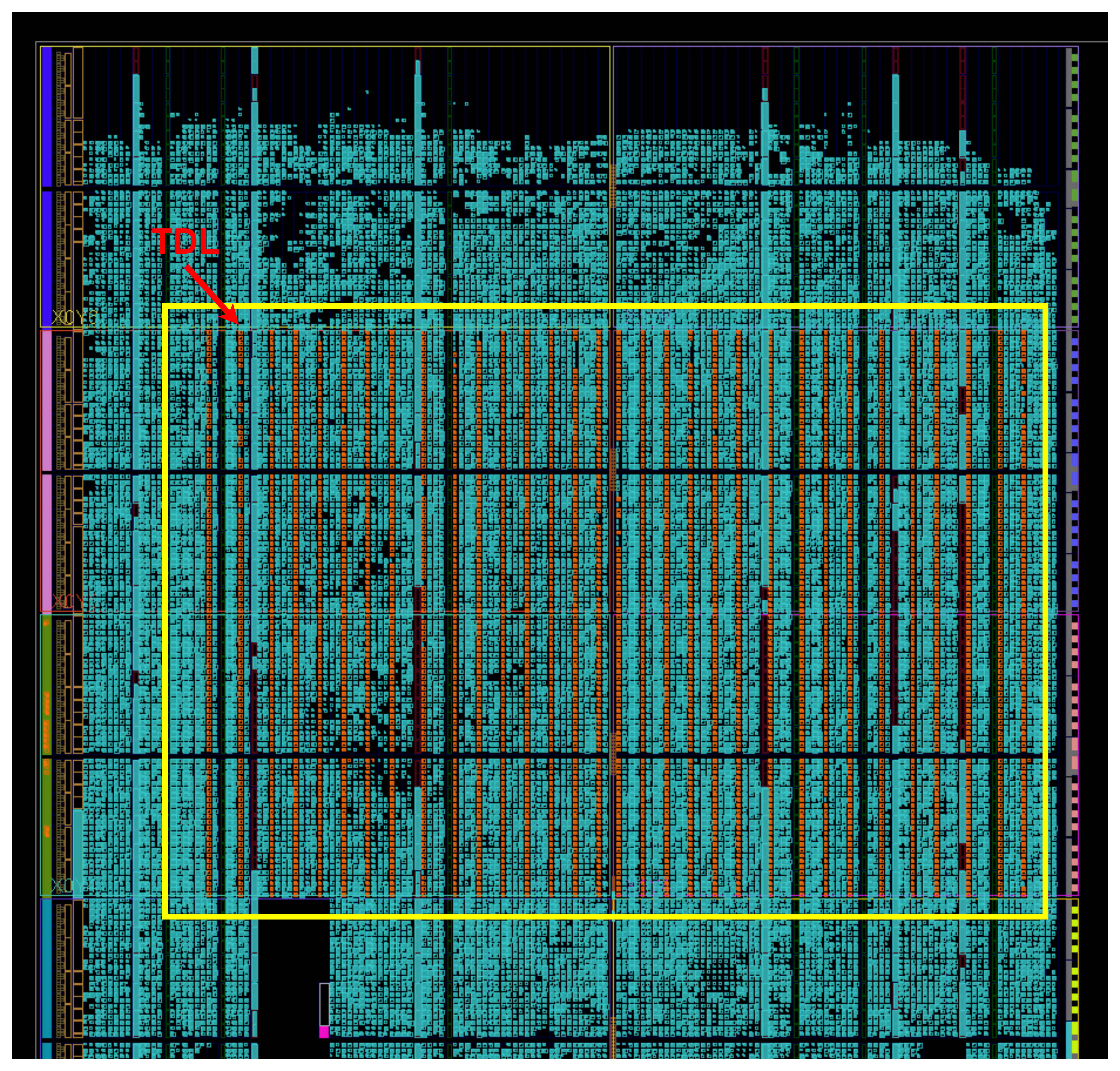}
    \caption{The 32-channel TDC with a severe bubble solution is implemented in a Xilinx Kintex-7 device and is located above the BUFG.}
    \label{fig:11}
\end{figure}

By combining the RMS, tap width distribution, and DNL of the two types of 2-channel TDCs, it can be concluded that the resolution and time precision are similar between the TDC located above the BUFG, which uses a severe bubble solution, and the TDC located below the BUFG, which does not use a severe bubble solution.  The TDC located above the BUFG with a severe bubble solution is expanded to 32 channels, and a coincidence time resolution test is conducted for each channel. The implementation of the 32-channel TDC with a severe bubble solution is shown in figure \ref{fig:11}.
\begin{figure}[htbp]
    \centering
    \includegraphics[width=0.8\linewidth]{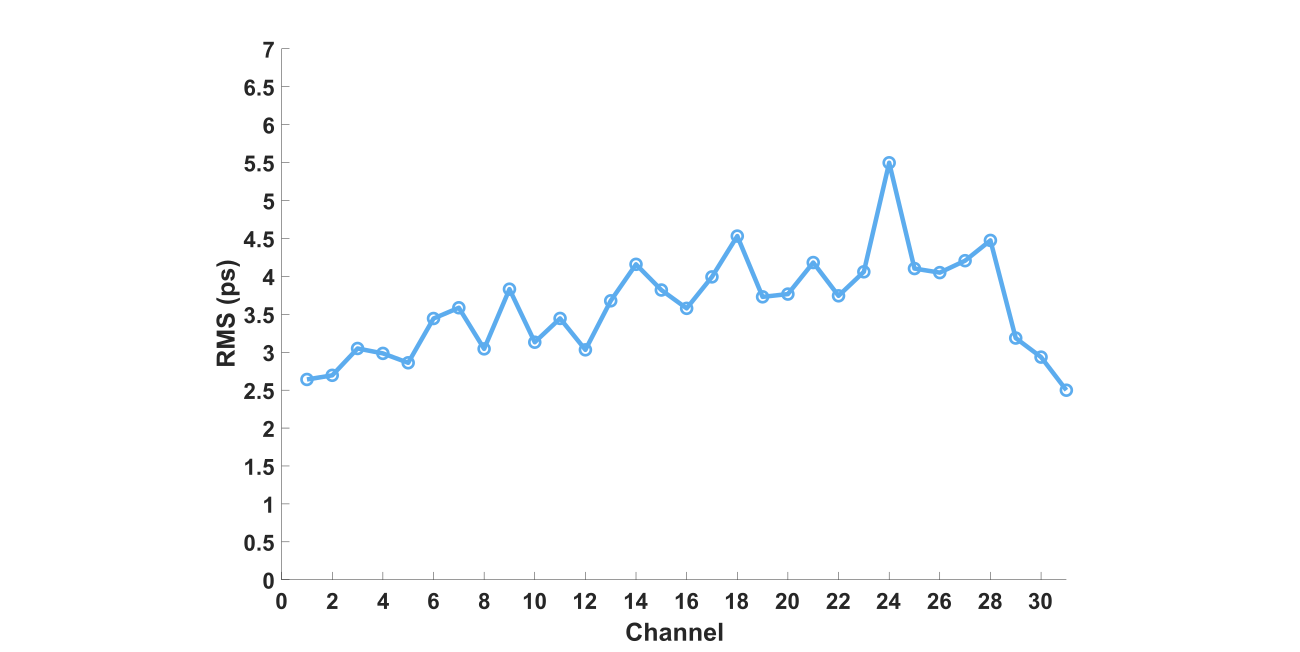}
    \caption{RMS time precision of the 32-channel TDC.}
    \label{fig:12}
\end{figure}

Channel \#0 of the 32-channel TDC performs coincidence time measurements with each of the other 31 TDC channels, and the testing environment is identical to that described for the 2-channel TDCs. The test results are shown in figure \ref{fig:12}, with a minimum RMS of approximately 2.50 ps, a maximum of approximately 5.50 ps, and an average of approximately 3.61 ps.

The average RMS measurement of the 32-channel TDC is not significantly different from that of the 2-channel TDC, which validates the severe bubble solution module. Furthermore, it also explains the feasibility of implementing multi-edge TDL TDC which is located above the BUFG in multi-channel schemes.

\subsection{Test of the 64-channel TDC}
\begin{figure}[htbp]
    \centering
    \begin{subfigure}{0.4\linewidth}
        \centering
        \includegraphics[width=\linewidth]{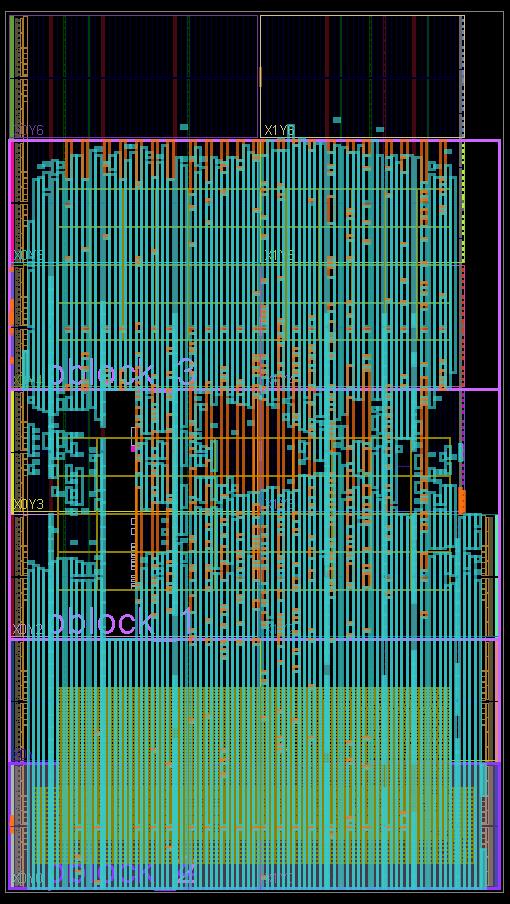}
        \caption{}
        \label{fig:13a}
    \end{subfigure}
    \begin{subfigure}{0.4\linewidth}
        \centering
        \includegraphics[width=\linewidth]{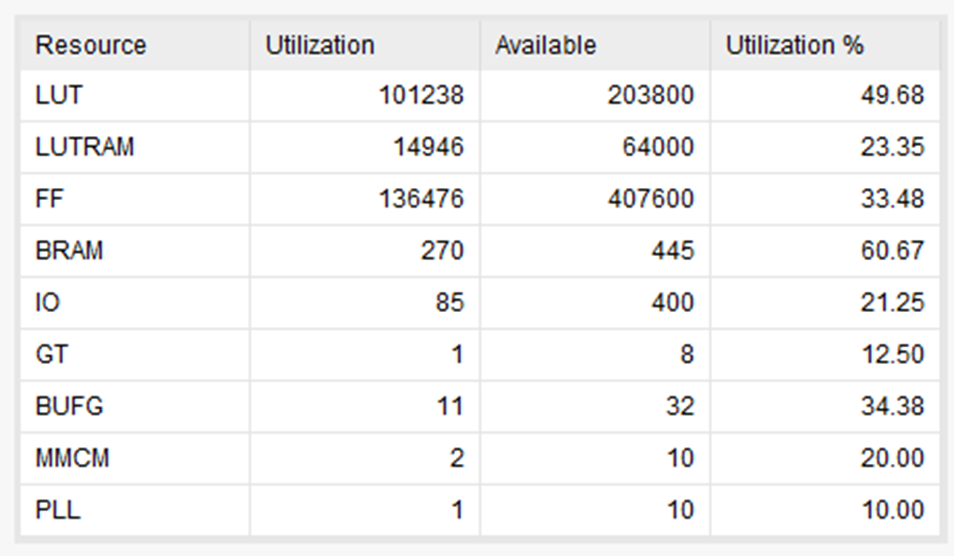}
        \caption{}
        \label{fig:13b}
    \end{subfigure}
    \caption{The 64-channel TDC implement in Xilinx Kintex-7 device. (a) The 64-channel TDC in FPGA device. (b) Resource usage of the 64-channel TDC in FPGA.}
    \label{fig:13}
\end{figure}
\begin{figure}[htbp]
    \centering
    \begin{subfigure}{0.49\linewidth}
        \centering
        \includegraphics[width=\linewidth]{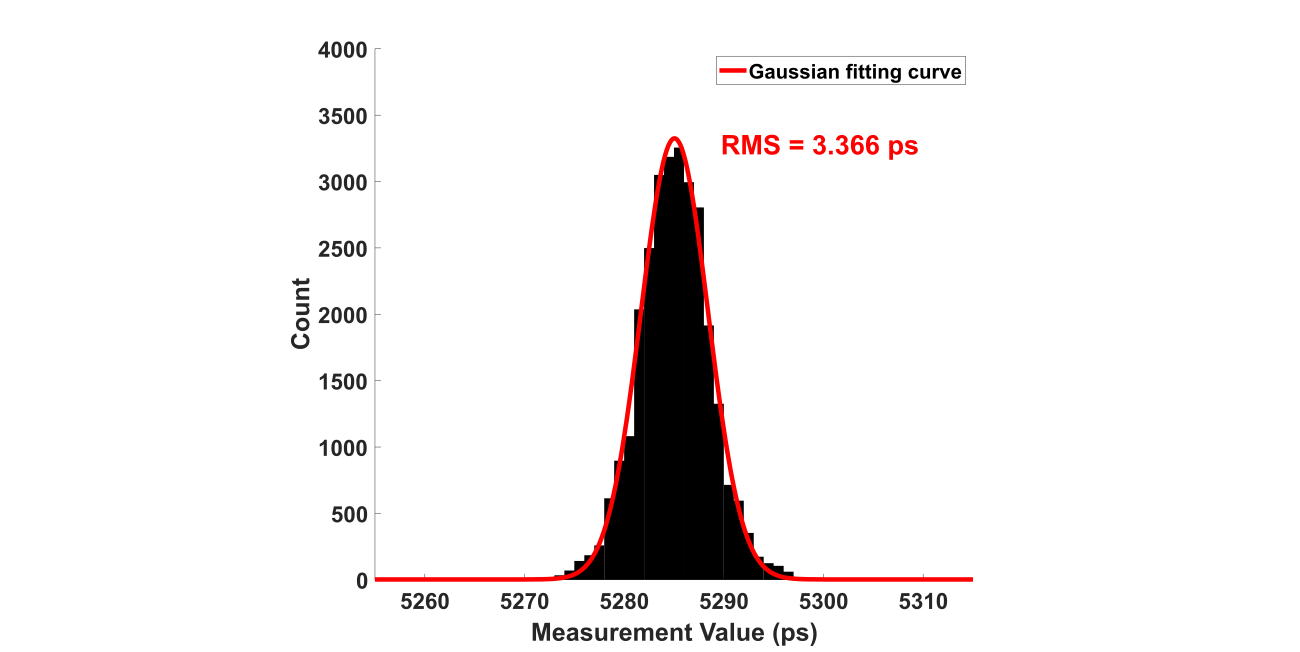}
        \caption{}
        \label{fig:14a}
    \end{subfigure}
    \begin{subfigure}{0.49\linewidth}
        \centering
        \includegraphics[width=\linewidth]{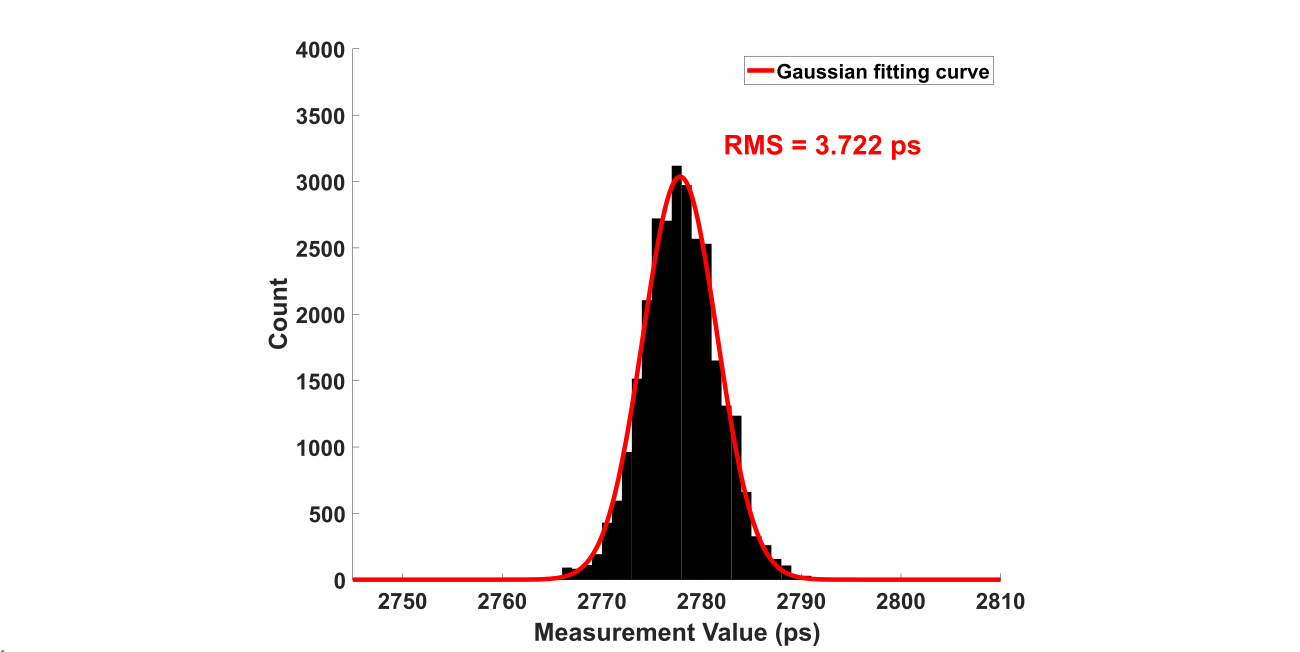}
        \caption{}
        \label{fig:14b}
    \end{subfigure}
    \caption{(a) DNL of the rising edge TDC \#0. (b) INL of the rising edge TDC \#0. (c) DNL of the falling edge TDC \#0. (d) INL of the falling edge TDC \#0.}
    \label{fig:14}
\end{figure}
\begin{figure}[htbp]
    \centering
    \includegraphics[width=0.7\linewidth]{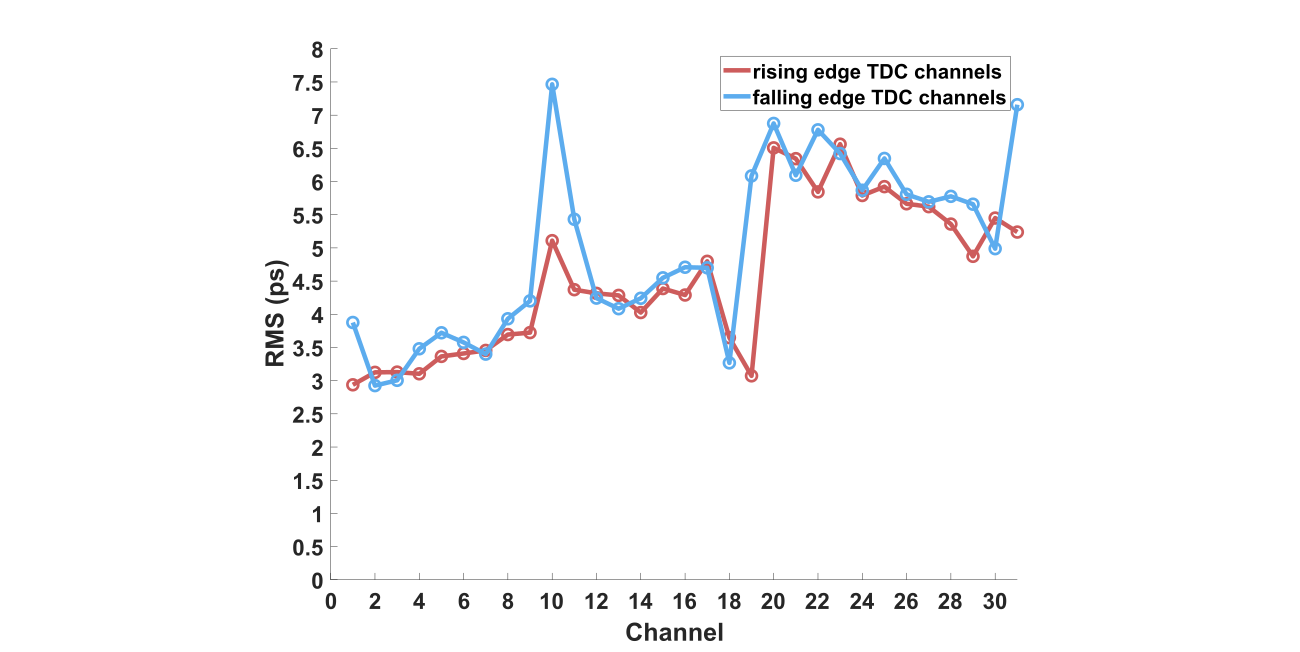}
    \caption{RMS time precision of the 64-channel TDC.}
    \label{fig:15}
\end{figure}

The implementation architecture of the 64-channel Time-to-Digital Converter (TDC) is illustrated in figure \ref{fig:13a}. It is divided into 32 channels dedicated to measuring the timing of rising edge and 32 channels dedicated to measuring the timing of falling edge. Figure \ref{fig:13b} shows the resource utilization rate of 64 channel TDC, with the highest resource utilization rate not exceeding 65\%.

The only distinction between falling edge channels and rising edge channels lies in the additional inverter at the input side. The channels are completely independent of each other, enabling the separate measurement of the sampling signal's rising and falling edges, and also allowing for the simultaneous measurement of both the rising and falling edges of same signal.

Testing environment being identical to that described in the Section IV.A, and the FPGA temperature was 69.3 °C. The sampling signal is fanned out into 64 routes within the FPGA, 32 of which are utilized for the measurement of rising edges, and the remaining 32 for the measurement of falling edges. The rising edge TDC \#0 performs coincidence time measurements with each of the other 31 rising edge TDC channels, while the falling edge TDC \#0 performs coincidence time measurements with each of the other 31 falling edge TDC channels. 

The coincidence time resolution testing results of the 64-channel TDC as shown in figure \ref{fig:15}. The coincidence time resolution testing results of the rising edge channel \#0 and channel \#5 as shown in figure \ref{fig:14a}, with the coincidence time resolution testing results of the falling edge channel \#0 and channel \#5 as shown in figure \ref{fig:14b}.

The minimum RMS for the rising edge TDC channels is approximately 2.94 ps, with a maximum of around 6.56 ps and an average of approximately 4.56 ps. For the falling edge TDC channels, the minimum RMS is approximately 2.93 ps, the maximum is approximately 7.46 ps, and the average is approximately 4.98 ps. The overall average RMS is around 4.77 ps, with a maximum deviation of 4.54 ps between channels. The uniformity of RMS precision among channels is relatively poor, probably relate to the clock crosstalk between TDC channels, or the implementation of multiple clock signals within the FPGA\cite{RN28}. Precision test based on temperature change was delayed because the temperature control box had not arrived yet.

\section{CONCLUSION}

Aiming at the serious bubble problem in TDL above the BUFG, a severe bubble solution based on tap swapping and ones-counter method is proposed. This strategy can process the severe bubble caused by clock region skew online without introducing new measurement errors, making it more convenient for multi-edge TDL TDC to be implemented in multi-channel scheme. 

A 64-channel 4-edge WU-A TDC has been implemented incorporating the multi-edge decomposition encoder within a 28 nm Xilinx Kintex-7 FPGA. The achieved 64-channel TDC exhibits an RMS precision of less than 8 ps, with an average RMS precision below 5 ps. This report also provides a detailed description of the implementation of the encoder. 

Experimental results demonstrate that the realized TDC meets the requirements of a multi-channel and high-precision, and is also suitable for other fields such as high-energy physics, biomedical imaging, general scientific instruments, etc.

\appendix

\acknowledgments

The authors are grateful to Dr.Yonggang Wang at the University of Science and Technology of China for sharing his design for one of the multi-edge wave union A TDC.


\begin{CJK}{UTF8}{gbsn}
\bibliography{mainref}

\begin{thebibliography}{10}
\expandafter\ifx\csname url\endcsname\relax
  \def\url#1{\texttt{#1}}\fi
\expandafter\ifx\csname urlprefix\endcsname\relax\def\urlprefix{URL }\fi
\expandafter\ifx\csname href\endcsname\relax
  \def\href#1#2{#2} \def\path#1{#1}\fi

\bibitem{RN1}
M.~P. Mattada, H.~Guhilot, Time-to-digital converters-a comprehensive review,
  INTERNATIONAL JOURNAL OF CIRCUIT THEORY AND APPLICATIONS 49~(3) (2021)
  778--800.
\newblock \href {https://doi.org/10.1002/cta.2936}
  {\path{doi:10.1002/cta.2936}}.

\bibitem{RN2}
刘树彬, 郭建华, 张艳丽, 赵龙, 安琪,
  高精度数据驱动型tdc在高能物理实验中应用的研究, 核技术
  29~(1) (2006) 72--76.

\bibitem{RN3}
W.~Yonggang, C.~Xinyi, L.~Deng, Z.~Wensong, L.~Chong, A linear
  time-over-threshold digitizing scheme and its 64-channel daq prototype design
  on fpga for a continuous crystal pet detector, IEEE Transactions on Nuclear
  Science 61~(1) (2014) 99--106.
\newblock \href {https://doi.org/10.1109/TNS.2013.2293758}
  {\path{doi:10.1109/TNS.2013.2293758}}.

\bibitem{RN4}
W.~S. Choong, Q.~Peng, C.~Q. Vu, B.~T. Turko, W.~W. Moses, High-performance
  electronics for time-of-flight pet systems, JOURNAL OF INSTRUMENTATION 8
  (2013).
\newblock \href {https://doi.org/10.1088/1748-0221/8/01/T01006}
  {\path{doi:10.1088/1748-0221/8/01/T01006}}.

\bibitem{RN5}
B.~A.~M. Zwaans, T.~Frach, B.~Zwaans, B.~A.~M. Zvaans, \href{<Go to
  ISI>://DIIDW:2012N04659}{High speed radiation detection method for positron
  emission tomography (pet), involves associating time stamp with event based
  on first and second time stamps, and comparison of time difference between
  time stamps and fixed time offset}.
\newline\urlprefix\url{<Go to ISI>://DIIDW:2012N04659}

\bibitem{RN6}
P.~Palojarvi, K.~Maatta, J.~Kostamovaara, Integrated time-of-flight laser
  radar, IEEE Transactions on Instrumentation and Measurement 46~(4) (1997)
  996--999.
\newblock \href {https://doi.org/10.1109/19.650815}
  {\path{doi:10.1109/19.650815}}.

\bibitem{RN7}
K.~Yoshioka, H.~Kubota, T.~Fukushima, S.~Kondo, T.~T. Ta, H.~Okuni,
  K.~Watanabe, M.~Hirono, Y.~Ojima, K.~Kimura, S.~Hosoda, Y.~Ota, T.~Koizumi,
  N.~Kawabe, Y.~Ishii, Y.~Iwagami, S.~Yagi, I.~Fujisawa, N.~Kano, T.~Sugimoto,
  D.~Kurose, N.~Waki, Y.~Higashi, T.~Nakamura, Y.~Nagashima, H.~Ishii, A.~Sai,
  N.~Matsumoto, A 20-ch tdc/adc hybrid architecture lidar soc for 240 $\times$
  96 pixel 200-m range imaging with smart accumulation technique and residue
  quantizing sar adc, IEEE Journal of Solid-State Circuits 53~(11) (2018)
  3026--3038.
\newblock \href {https://doi.org/10.1109/JSSC.2018.2868315}
  {\path{doi:10.1109/JSSC.2018.2868315}}.

\bibitem{RN8}
P.~Vines, K.~Kuzmenko, J.~Kirdoda, D.~C.~S. Dumas, M.~M. Mirza, R.~W. Millar,
  D.~J. Paul, G.~S. Buller,
  \href{https://doi.org/10.1038/s41467-019-08830-w}{High performance planar
  germanium-on-silicon single-photon avalanche diode detectors}, Nature
  Communications 10~(1) (2019) 1086.
\newblock \href {https://doi.org/10.1038/s41467-019-08830-w}
  {\path{doi:10.1038/s41467-019-08830-w}}.
\newline\urlprefix\url{https://doi.org/10.1038/s41467-019-08830-w}

\bibitem{RN9}
I.~Rausch, A.~Ruiz, I.~Valverde-Pascual, J.~Cal-González, T.~Beyer, I.~Carrio,
  \href{http://jnm.snmjournals.org/content/60/4/561.abstract}{Performance
  evaluation of the vereos pet/ct system according to the nema nu2-2012
  standard}, Journal of Nuclear Medicine 60~(4) (2019) 561.
\newblock \href {https://doi.org/10.2967/jnumed.118.215541}
  {\path{doi:10.2967/jnumed.118.215541}}.
\newline\urlprefix\url{http://jnm.snmjournals.org/content/60/4/561.abstract}

\bibitem{RN10}
J.~van Sluis, J.~de~Jong, J.~Schaar, W.~Noordzij, P.~van Snick, R.~Dierckx,
  R.~Borra, A.~Willemsen, R.~Boellaard, Performance characteristics of the
  digital biograph vision pet/ct system, J Nucl Med 60~(7) (2019) 1031--1036.
\newblock \href {https://doi.org/10.2967/jnumed.118.215418}
  {\path{doi:10.2967/jnumed.118.215418}}.

\bibitem{RN11}
L.~Ma, L.~Chen, P.~Chai, Z.~Liang, G.~Huang, J.~Hu, X.~Han, Z.~Hua, X.~Huang,
  M.~Jin, X.~Jiang, Z.~Jin, S.~Liu, W.~Pan, S.~Qian, L.~Ren, S.~Si, J.~Sun,
  L.~Wei, Q.~Wu, T.~Wang, X.~Wang, Y.~Wang, Y.~Wang, Z.~Wang, Z.~Wang, N.~Wang,
  K.~Wu, X.~Yan, H.~Zhang, Z.~Zhang,
  \href{https://www.sciencedirect.com/science/article/pii/S0168900223000797}{A
  novel multi-anode mcp-pmt with cherenkov radiator window}, Nuclear
  Instruments and Methods in Physics Research Section A: Accelerators,
  Spectrometers, Detectors and Associated Equipment 1049 (2023) 168089.
\newblock \href {https://doi.org/https://doi.org/10.1016/j.nima.2023.168089}
  {\path{doi:https://doi.org/10.1016/j.nima.2023.168089}}.
\newline\urlprefix\url{https://www.sciencedirect.com/science/article/pii/S0168900223000797}

\bibitem{RN12}
H.~Chen, D.~D.-U. Li, \href{<Go to ISI>://WOS:000452731500075}{Multichannel,
  low nonlinearity time-to-digital converters based on 20 and 28 nm fpgas},
  Ieee Transactions on Industrial Electronics 66~(4) (2019) 3265--3274.
\newblock \href {https://doi.org/10.1109/tie.2018.2842787}
  {\path{doi:10.1109/tie.2018.2842787}}.
\newline\urlprefix\url{<Go to ISI>://WOS:000452731500075}

\bibitem{RN13}
Y.~Wang, Q.~Cao, C.~Liu, A multi-chain merged tapped delay line for high
  precision time-to-digital converters in fpgas, IEEE Transactions on Circuits
  and Systems II: Express Briefs 65~(1) (2018) 96--100.
\newblock \href {https://doi.org/10.1109/TCSII.2017.2698479}
  {\path{doi:10.1109/TCSII.2017.2698479}}.

\bibitem{RN14}
P.~Chen, Y.~Y. Hsiao, Y.~S. Chung, W.~X. Tsai, J.~M. Lin, A 2.5-ps bin size and
  6.7-ps resolution fpga time-to-digital converter based on delay wrapping and
  averaging, IEEE Transactions on Very Large Scale Integration (VLSI) Systems
  25~(1) (2017) 114--124.
\newblock \href {https://doi.org/10.1109/TVLSI.2016.2569626}
  {\path{doi:10.1109/TVLSI.2016.2569626}}.

\bibitem{RN15}
T.~Sui, Z.~Zhao, S.~Xie, Y.~Xie, Y.~Zhao, Q.~Huang, J.~Xu, Q.~Peng, A 2.3-ps
  rms resolution time-to-digital converter implemented in a low-cost cyclone v
  fpga, IEEE Transactions on Instrumentation and Measurement 68~(10) (2019)
  3647--3660.
\newblock \href {https://doi.org/10.1109/TIM.2018.2880940}
  {\path{doi:10.1109/TIM.2018.2880940}}.

\bibitem{RN16}
R.~Szplet, D.~Sondej, G.~Grzeda, Subpicosecond-resolution time-to-digital
  converter with multi-edge coding in independent coding lines, in: 2014 IEEE
  International Instrumentation and Measurement Technology Conference (I2MTC)
  Proceedings, pp. 747--751.
\newblock \href {https://doi.org/10.1109/I2MTC.2014.6860842}
  {\path{doi:10.1109/I2MTC.2014.6860842}}.

\bibitem{RN17}
Y.~Wang, W.~Xie, H.~Chen, D.~Day-Uei~Li,
  \href{https://www.sciencedirect.com/science/article/pii/S0263224122014543}{High-resolution
  time-to-digital converters (tdcs) with a bidirectional encoder}, Measurement
  206 (2023) 112258.
\newblock \href
  {https://doi.org/https://doi.org/10.1016/j.measurement.2022.112258}
  {\path{doi:https://doi.org/10.1016/j.measurement.2022.112258}}.
\newline\urlprefix\url{https://www.sciencedirect.com/science/article/pii/S0263224122014543}

\bibitem{RN18}
R.~Machado, J.~Cabral, F.~S. Alves, Recent developments and challenges in
  fpga-based time-to-digital converters, IEEE Transactions on Instrumentation
  and Measurement 68~(11) (2019) 4205--4221.
\newblock \href {https://doi.org/10.1109/TIM.2019.2938436}
  {\path{doi:10.1109/TIM.2019.2938436}}.

\bibitem{RN19}
X.~Mao, F.~Yang, F.~Wei, J.~Shi, J.~Cai, H.~Cai,
  \href{https://www.mdpi.com/1424-8220/22/6/2306}{A low temperature coefficient
  time-to-digital converter with 1.3 ps resolution implemented in a 28 nm
  fpga}, Sensors 22~(6) (2022) 2306.
\newline\urlprefix\url{https://www.mdpi.com/1424-8220/22/6/2306}

\bibitem{RN20}
J.~Wu, Z.~Shi, \href{https://ieeexplore.ieee.org/document/4775079/}{The 10-ps
  wave union tdc: Improving fpga tdc resolution beyond its cell delay}, in:
  2008 IEEE Nuclear Science Symposium Conference Record, pp. 3440--3446.
\newblock \href {https://doi.org/10.1109/NSSMIC.2008.4775079}
  {\path{doi:10.1109/NSSMIC.2008.4775079}}.
\newline\urlprefix\url{https://ieeexplore.ieee.org/document/4775079/}

\bibitem{RN21}
J.~Kuang, Y.~Wang, Q.~Cao, C.~Liu,
  \href{https://www.sciencedirect.com/science/article/pii/S0168900218302183}{Implementation
  of a high precision multi-measurement time-to-digital convertor on a kintex-7
  fpga}, Nuclear Instruments and Methods in Physics Research Section A:
  Accelerators, Spectrometers, Detectors and Associated Equipment 891 (2018)
  37--41.
\newblock \href {https://doi.org/https://doi.org/10.1016/j.nima.2018.02.064}
  {\path{doi:https://doi.org/10.1016/j.nima.2018.02.064}}.
\newline\urlprefix\url{https://www.sciencedirect.com/science/article/pii/S0168900218302183}

\bibitem{RN22}
Y.~Wang, J.~Kuang, C.~Liu, Q.~Cao, A 3.9-ps rms precision time-to-digital
  converter using ones-counter encoding scheme in a kintex-7 fpga, IEEE
  Transactions on Nuclear Science 64~(10) (2017) 2713--2718.
\newblock \href {https://doi.org/10.1109/TNS.2017.2746626}
  {\path{doi:10.1109/TNS.2017.2746626}}.

\bibitem{RN23}
S.~Bourdeauducq, A 26 ps rms time-to-digital converter core for spartan-6 fpgas
  (03 2013).
\newblock \href {https://doi.org/10.48550/arXiv.1303.6840}
  {\path{doi:10.48550/arXiv.1303.6840}}.

\bibitem{RN24}
Y.~Wang, X.~Zhou, Z.~Song, J.~Kuang, Q.~Cao, A 3.0-ps rms precision
  277-msamples/s throughput time-to-digital converter using multi-edge encoding
  scheme in a kintex-7 fpga, IEEE Transactions on Nuclear Science 66~(10)
  (2019) 2275--2281.
\newblock \href {https://doi.org/10.1109/TNS.2019.2938571}
  {\path{doi:10.1109/TNS.2019.2938571}}.

\bibitem{RN25}
X.~Deng, Q.~Chen, \href{https://dx.doi.org/10.1088/1748-0221/16/12/P12031}{A
  4.32-ps precision time-to-digital convertor using multisampling wave union
  method on a 28-nm fpga}, Journal of Instrumentation 16~(12) (2021) P12031.
\newblock \href {https://doi.org/10.1088/1748-0221/16/12/P12031}
  {\path{doi:10.1088/1748-0221/16/12/P12031}}.
\newline\urlprefix\url{https://dx.doi.org/10.1088/1748-0221/16/12/P12031}

\bibitem{RN26}
P.~Kwiatkowski, R.~Szplet, Efficient implementation of multiple time coding
  lines-based tdc in an fpga device, IEEE Transactions on Instrumentation and
  Measurement 69~(10) (2020) 7353--7364.
\newblock \href {https://doi.org/10.1109/TIM.2020.2984929}
  {\path{doi:10.1109/TIM.2020.2984929}}.

\bibitem{RN27}
H.~Menninga, C.~Favi, M.~W. Fishburn, E.~Charbon, A multi-channel, 10ps
  resolution, fpga-based tdc with 300ms/s throughput for open-source pet
  applications, in: 2011 IEEE Nuclear Science Symposium Conference Record, pp.
  1515--1522.
\newblock \href {https://doi.org/10.1109/NSSMIC.2011.6154362}
  {\path{doi:10.1109/NSSMIC.2011.6154362}}.

\bibitem{RN28}
C.~Liu, Y.~Wang, \href{<Go to ISI>://WOS:000356456100025}{A 128-channel, 710 m
  samples/second, and less than 10 ps rms resolution time-to-digital converter
  implemented in a kintex-7 fpga}, Ieee Transactions on Nuclear Science 62~(3)
  (2015) 773--783.
\newblock \href {https://doi.org/10.1109/tns.2015.2421319}
  {\path{doi:10.1109/tns.2015.2421319}}.
\newline\urlprefix\url{<Go to ISI>://WOS:000356456100025}

\end{thebibliography}
\end{CJK}



\end{document}